\documentclass{aa}  

\usepackage{placeins}

\usepackage{amsmath}
\usepackage{graphicx}
\usepackage{adjustbox}
\def\be{\begin{equation}}
\def\ee{\end{equation}}

\usepackage{xcolor}

\newcommand{\srg}{\textit{SRG}}

\DeclareUnicodeCharacter{2212}{\textendash}

\usepackage{graphicx}	
\usepackage{comment}

\def\lsim{\;\raise0.3ex\hbox{$<$\kern-0.75em\raise-1.1ex\hbox{$\sim$}}\;}
\def\gsim{\;\raise0.3ex\hbox{$>$\kern-0.75em\raise-1.1ex\hbox{$\sim$}}\;}
\newcommand{\msun}{\,$M_{\odot}$}
\newcommand{\ergs}{\,erg\,s$^{-1}$\,}
\newcommand{\kms}{\,km\,s$^{-1}$}

\newcommand{\gcmq}{\,g\,cm$^{-3}$}

\newcommand{\cmq}{\,cm$^{-3}$}
\newcommand{\ha}{H$\alpha$\,}
\newcommand{\msyr}{\,$M_{\odot}\,{\rm yr^{-1}}$}

\newcommand{\psr}{PSR~J0538+2817\,}

\usepackage{hyperref}

\begin{document}

\title{
Study of X-ray emission from the S147 nebula by \textit{SRG}/eROSITA: supernova-in-the-cavity scenario
}
\author{
Ildar~I.~Khabibullin \inst{1,2,3} 
\and
Eugene~M.~Churazov \inst{2,3} 
\and
Nikolai~N.~Chugai \inst{4} 
\and
Andrei~M.~Bykov \inst{5} 
\and
Rashid~A.~Sunyaev \inst{2,3} 
\and
Victor~P.~Utrobin \inst{6,4} 
\and
Igor~I.~Zinchenko \inst{7}
\and
Miltiadis Michailidis\inst{8}
\and
Gerd P{\"u}hlhofer\inst{8}
\and
Werner Becker\inst{9,10}
\and
Michael Freyberg\inst{9}
\and
Andrea Merloni\inst{9}
\and
Andrea Santangelo\inst{8}
\and
Manami Sasaki\inst{11}
}
\institute{Universitäts-Sternwarte, Fakultät für Physik, Ludwig-Maximilians-Universität München, Scheinerstr.1, 81679 München, Germany
\and
Space Research Institute (IKI), Profsoyuznaya 84/32, Moscow 117997, Russia
\and
Max Planck Institute for Astrophysics, Karl-Schwarzschild-Str. 1, D-85741 Garching, Germany 
\and 
Institute of Astronomy, Russian Academy of Sciences, 48 Pyatnitskaya str., Moscow 119017, Russia
\and
Ioffe Institute, Politekhnicheskaya st. 26, Saint Petersburg 194021, Russia
\and
NRC `Kurchatov Institute', acad. Kurchatov Square 1, Moscow 123182, Russia
\and
Institute of Applied Physics of the Russian Academy of Sciences, 46 Ul'yanov~str., Nizhny Novgorod 603950, Russia
 \and 
 Institut für Astronomie und Astrophysik Tübingen (IAAT), Sand 1, 72076 Tübingen, Germany
\and
Max-Planck Institut für extraterrestrische Physik, Giessenbachstraße, 85748 Garching, Germany
 \and 
 Max-Planck Institut für Radioastronomie, Auf dem Hügel 69, 53121 Bonn, Germany
 \and 
 Dr.\ Karl Remeis Observatory, Erlangen Centre for Astroparticle Physics, Friedrich-Alexander-Universit\"{a}t Erlangen-N\"{u}rnberg, Sternwartstra{\ss}e 7, 96049 Bamberg, Germany 
}


\abstract{
The Simeis~147 nebula (S147), particularly well known for a spectacular net of ${\rm H}_\alpha$-emitting filaments, is often considered one of the largest and oldest known supernova remnants in the Milky Way. Here, and in a companion paper, we present studies of X-ray emission from the S147 nebula using the data of SRG/eROSITA All-Sky Survey observations. In this paper, we argue that  many inferred properties of the X-ray emitting gas are broadly consistent with a scenario of the supernova explosion in a low-density cavity, e.g. a wind-blown-bubble. This scenario assumes that a $\sim 20\,{\rm M_\odot}$ progenitor star has had small velocity with respect to the ambient ISM, so it stayed close to the center of a dense shell created during its Main Sequence evolution till the moment of the core-collapse explosion. The ejecta first propagate through the low-density cavity until they collide with the dense shell, and only then the reverse shock goes deeper into the ejecta and powers the observed X-ray emission of the nebula. The part of the remnant inside the dense shell remains non-radiative till now and, plausibly, in a state with $T_e<T_i$ and Non-Equilibrium Ionization (NEI). On the contrary, the forward shock becomes radiative immediately after entering the dense shell, and, being subject to instabilities, creates a characteristic "foamy" appearance of the nebula in ${\rm H}_\alpha$ and radio emission. 
}


\titlerunning{SN-in-a-cavity scenario for S147}


   \maketitle

\section{Introduction}
\label{sec:intro}

Massive stars, although relatively rare, are capable of strongly affecting and shaping the surrounding interstellar medium (ISM) via energetic winds launched during different phases of their evolution, as well as via blast waves of the core-collapse explosion marking the ends of their lives \citep[e.g.][]{2023Galax..11...78D}. As was realized and described early on, the evolution of the supernova blast wave inside a wind-blown cavity differs substantially from the self-similar solution of a point explosion in a uniform medium, being characterized by a prolonged phase of the free expansion (until ejecta hit the walls of the cavity) and rapid onset of the radiative phase of the forward shock launched into the dense shell \citep{1989ApJ...344..332C,1991MNRAS.251..318T}. Although such a picture should be relevant for the bulk of the massive stars, the relative motion of the star and surrounding ISM apparently makes in-cavity explosions rare. 

Simeis 147 nebula (hereafter S147 for brevity) discovered  by G.~A.~Shajn  \citep{1952IzKry...9...52G} is commonly referred to as ``Spaghetti Nebula'' due to its spectacular filamentary appearance in H$_\alpha$ line emission \citep[][]{1976AZh....53...38L}. Non-thermal radio \citep[][]{1974AJ.....79.1253D,1980PASJ...32....1S,1986A&A...163..185F,2008A&A...482..783X,2024MNRAS.527.5683K} and gamma-ray emission \citep[][]{2012ApJ...752..135K,Suzuki22} has been detected from it, indicating that it is most likely a remnant of a supernova explosion (SNR~G180.0-01.7), possibly interacting with a molecular cloud at one of its boundaries \citep[see also ][ for an update]{Miltos2024}. A relatively old radio and X-ray pulsar was discovered within the extent of the nebula\citep[][]{1996ApJ...468L..55A,2003ApJ...593L..31K}, for which the measured proper motion directs away from the geometrical center of the nebula \citep{Romani_2003}, indicating that S147 might be indeed powered by a core-collapse supernova explosion $\sim30,000$ yrs ago \citep[][]{2006A&A...454..239G,2007ApJ...654..487N}.

The size of S147 is however indicative of a much older explosion,$\sim150,000$ yrs, if canonical values of the explosion energy and the density of the surrounding medium are assumed. This "age dilemma" \citep{Reich_2003,Romani_2003} can be reconciled if a scenario of an explosion in the wind-blown cavity is invoked, allowing the size of the object to be determined by winds before the explosion \citep{Reich_2003,2006A&A...454..239G}. One of the predictions in this picture is that after the passage of the reverse shock through the supernova ejecta, the cavity should be filled with the hot X-ray emitting gas, bearing traces of the enrichment by explosive nucleosynthesis products.  

Given the large size of S147, this emission is expected to be rather faint and difficult to observe with focusing X-ray telescopes featuring small Field-of-View, like \textit{Chandra}, \textit{XMM-Newton} or \textit{Suzaku}. In the course of the all-sky survey, a map of the full extent of the nebula was obtained by \srg/eROSITA, resulting in the first clear detection of soft thermal X-ray emission from S147 with most of the flux coming at energies below $\sim 1\,{\rm keV}$. 

In this paper, we report some of the key findings regarding spatial and spectral properties of the X-ray emission detected from the Spaghetti Nebula and discuss them in relation to the physical scenario of a supernova explosion in an interstellar cavity created by the progenitor star during its main sequence phase. A comprehensive description of the data analysis and their comparison to the data at other wavelengths is given in an accompanying paper \citep{Miltos2024}.

The paper is structured as follows: we describe X-ray observations in Section \ref{sec:obs} and outline morphological and spectral properties of the newly discovered S147 X-ray emission in Section~\ref{sec:xrays}. The model of a supernova explosion in a wind-blown bubble, capable of explaining the major properties of the object is presented in Section~\ref{sec:model} and discussed in Section~\ref{sec:discussion}. Conclusions are summarized in Section~\ref{sec:conclusions}  

%

\section{Observations}
\label{sec:obs}

The \srg\, X-ray observatory \citep{2021A&A...656A.132S}  was launched from the Baikonur cosmodrome on July 13, 2019. It carries two wide-angle grazing-incidence X-ray telescopes, eROSITA \citep{2021A&A...647A...1P} and the Mikhail Pavlinsky ART-XC telescope \citep{2021A&A...650A..42P}, which operate in the overlapping energy bands of 0.2–8 and 4–30 keV, respectively. The main dataset used here includes the data of \srg~/eROSITA from \textit{all four} consecutive all-sky surveys.

For imaging analysis, the data of all seven eROSITA telescope modules were used, while the data of two telescope modules (TM) without the on-chip optical blocking filter were excluded from the spectral analysis due to their different spectral response function and susceptibility to optical light leak contamination at low energies \citep[e.g.][]{2021A&A...647A...1P}.  

\section{X-ray emission}
\label{sec:xrays}

A comprehensive description of various properties of X-ray emission detected in the direction of the S147 nebula is presented in the companion paper, here we focus on 
key features 
that might 
support
the physical scenario we put forward in Section \ref{sec:model}.

\subsection{Broad-band X-ray morphology}

\begin{figure}
    \centering
    \includegraphics[angle=0,bb=50 220 510 640,width=0.99\columnwidth]{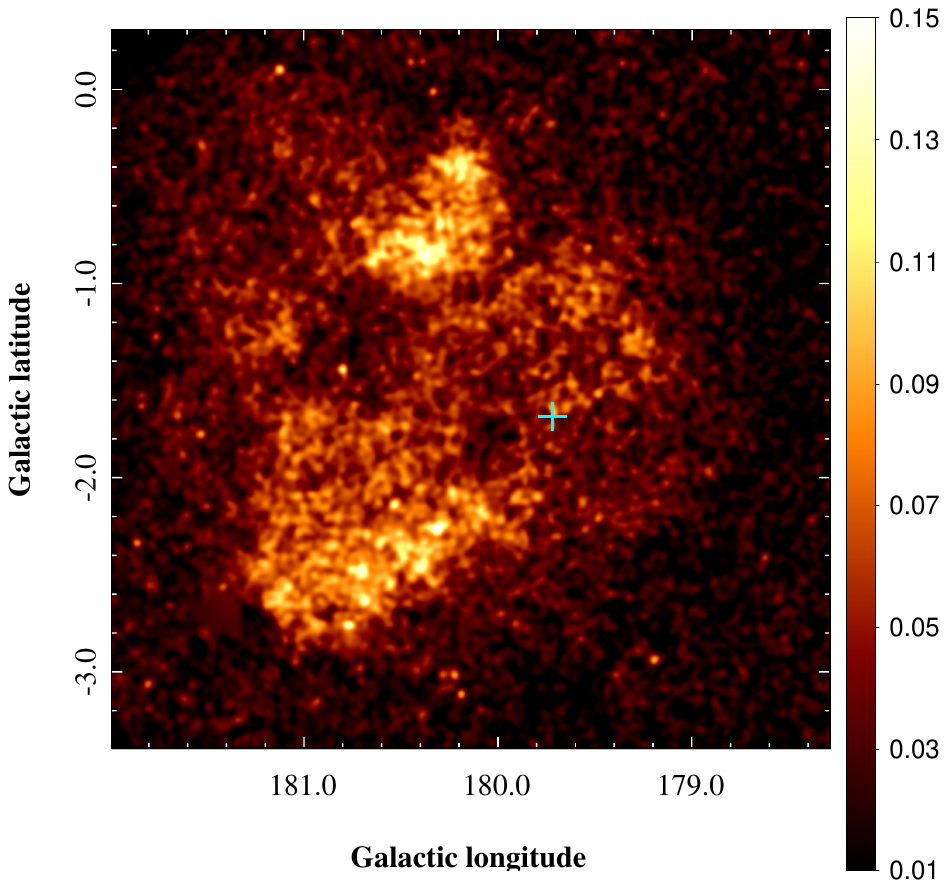}
    \caption{Broad-band (0.5-1.0 keV) (particle) background-subtracted exposure-corrected X-ray image (linear scale) obtained by \srg/eROSITA in the S147 direction (3.7$^\circ\times$3.7$^\circ$ in Galactic coordinates) after masking of point and mildly extended sources and smoothing with a Gaussian kernel with $\sigma=1'$.  This band maximizes the source-to-background ratio for the SNR emission. The cross marks the position of the pulsar \psr.}
    \label{fig:xraybroad}
\end{figure}

Figure \ref{fig:xraybroad} shows surface brightness of the X-ray emission in 0.5-1 keV band from a 3.7$^\circ$x3.7$^\circ$ patch (in Galactic coordinates) covering the full extent of S147. The image was produced by masking the detected point and mildly extended sources and smooth with a Gaussian kernel with $\sigma=1'$. The exact procedure is identical to the one used in the previous works on the newly discovered X-ray supernova remnants \citep{2021MNRAS.507..971C,2023MNRAS.521.5536K} and is described in the companion paper \citep{Miltos2024}. The image is centred close to the geometrical center of the H$\alpha$ and radio emission, at $(l,b)\approx (180.32^\circ,-1.65^\circ)$ \citep[e.g.][]{2003ApJ...593L..31K}, and the position of the pulsar \psr marked with the cross (the point-like emission of the pulsar itself has been masked). The ratio of the X-ray emission from the SNR to the unrelated background and foreground emission is maximized in this band (the latter is estimated from the adjacent sky regions outside the extent of the nebula). At lower and higher energies, the S147 emission falls below the background level making it barely visible in the 2D image, but still detectable in the background-subtracted source spectrum.

\begin{figure}
\centering
\includegraphics[angle=0,trim=1cm 7cm 2cm 4cm,width=0.95\columnwidth]{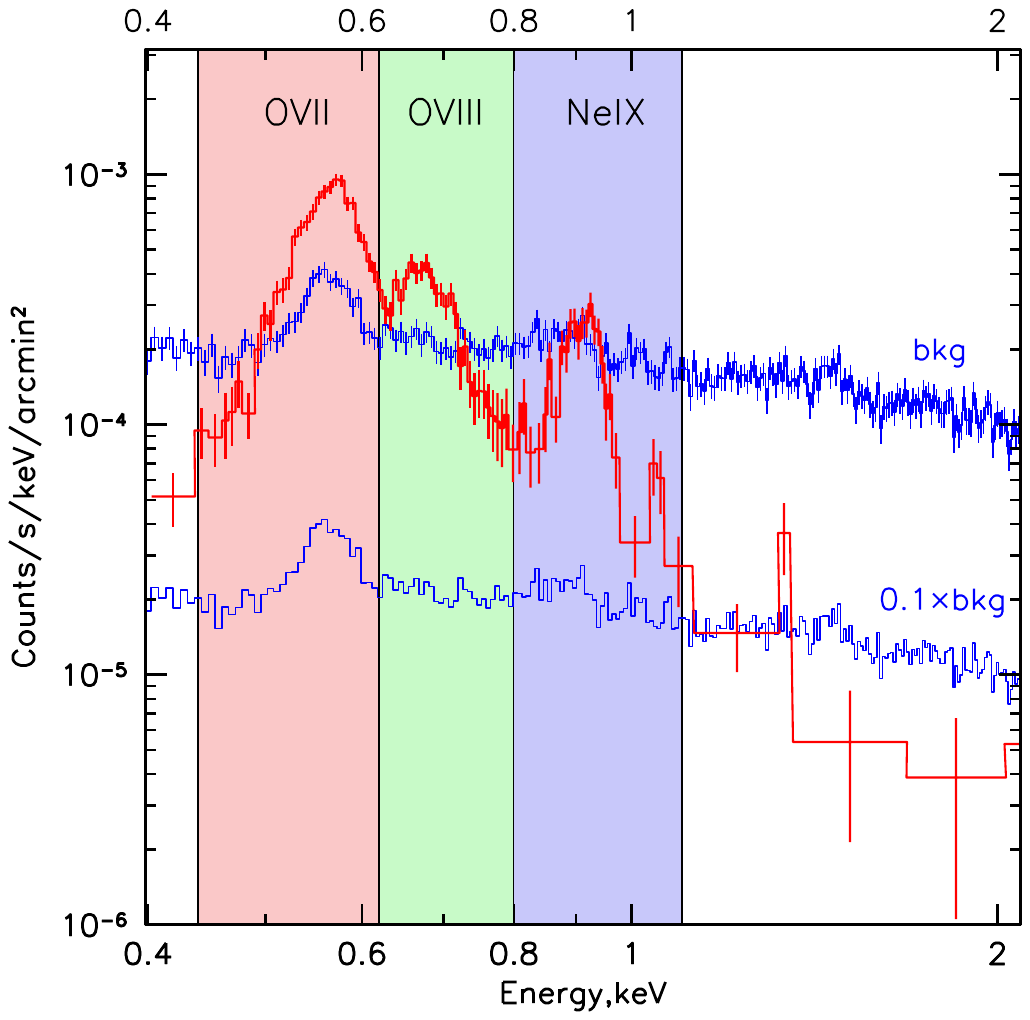}
\caption{{X-ray spectrum of the whole remnant (red data points) after subtraction of the background signal (blue data points) estimated from an adjacent sky region. Also shown is the level of 10\% of the background emission aimed at showing that above 1.5 keV the supernova signal amounts to a few \% of the background level, making conclusions regarding its spectral shape at these energies strongly background-sensitive. The three bands containing the brightest emission lines are shown in red (O~VII), green (O~VII), and blue (Ne~IX), and are used for RGB composite images.}}  
\label{fig:spec_bkg}
\end{figure}

\begin{figure}
    \centering
    \includegraphics[angle=0,bb=50 150 550 670,width=0.8\columnwidth]{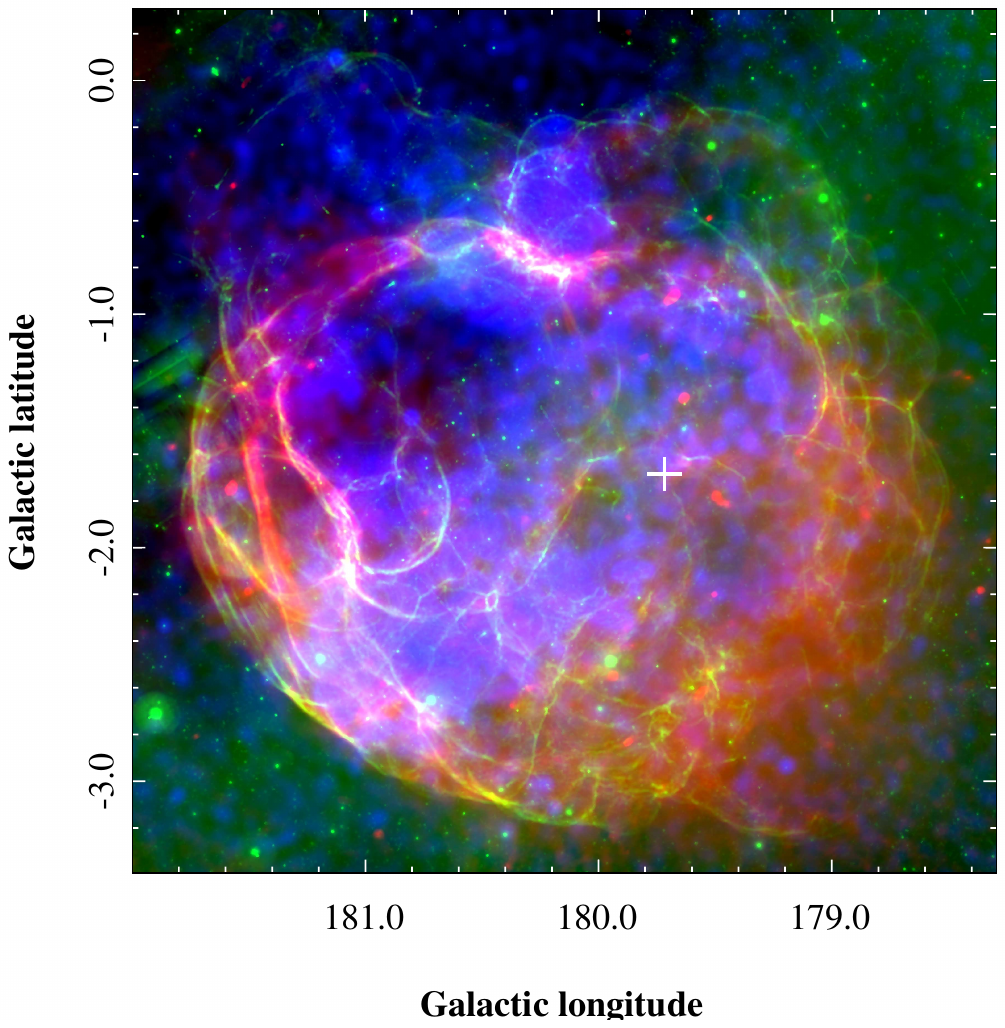}
      \includegraphics[angle=0,bb=60 170 540 645,width=0.47\columnwidth]{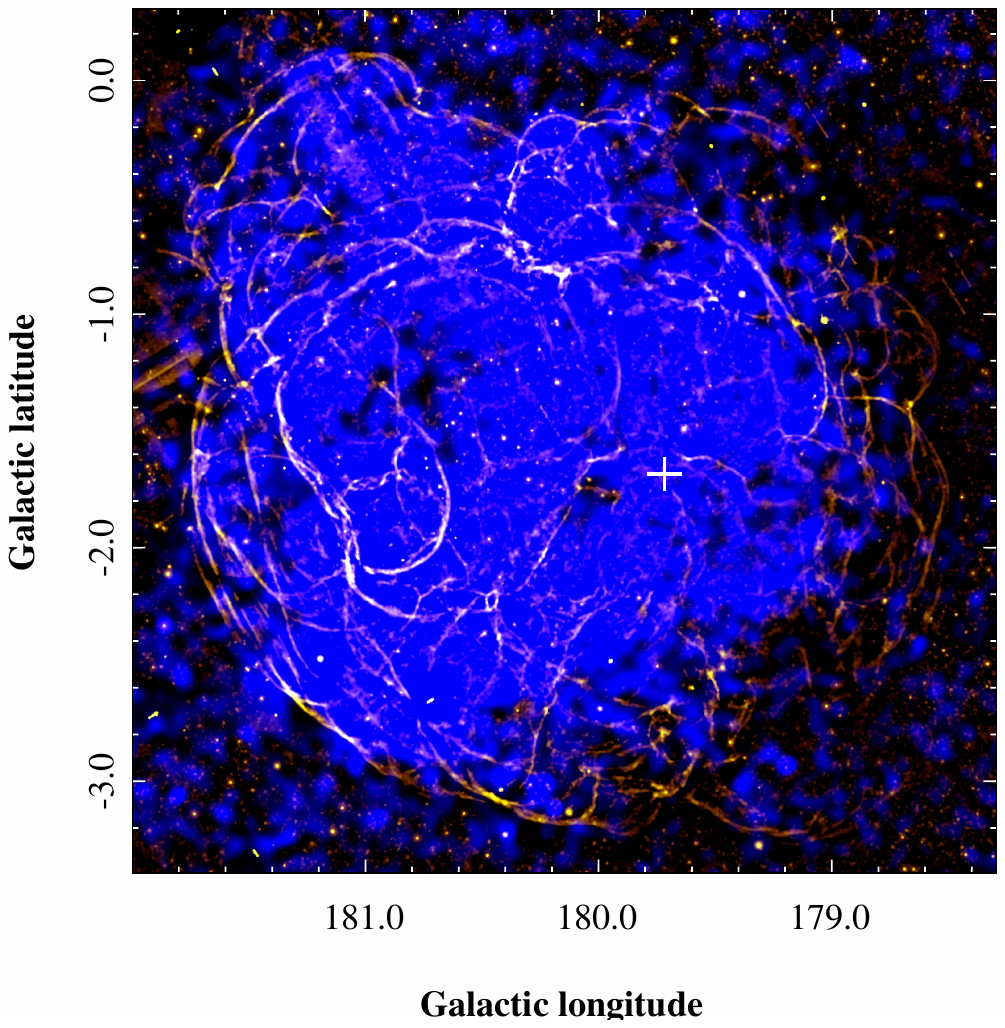}
    \includegraphics[angle=0,bb=60 170 540 646,width=0.47\columnwidth]{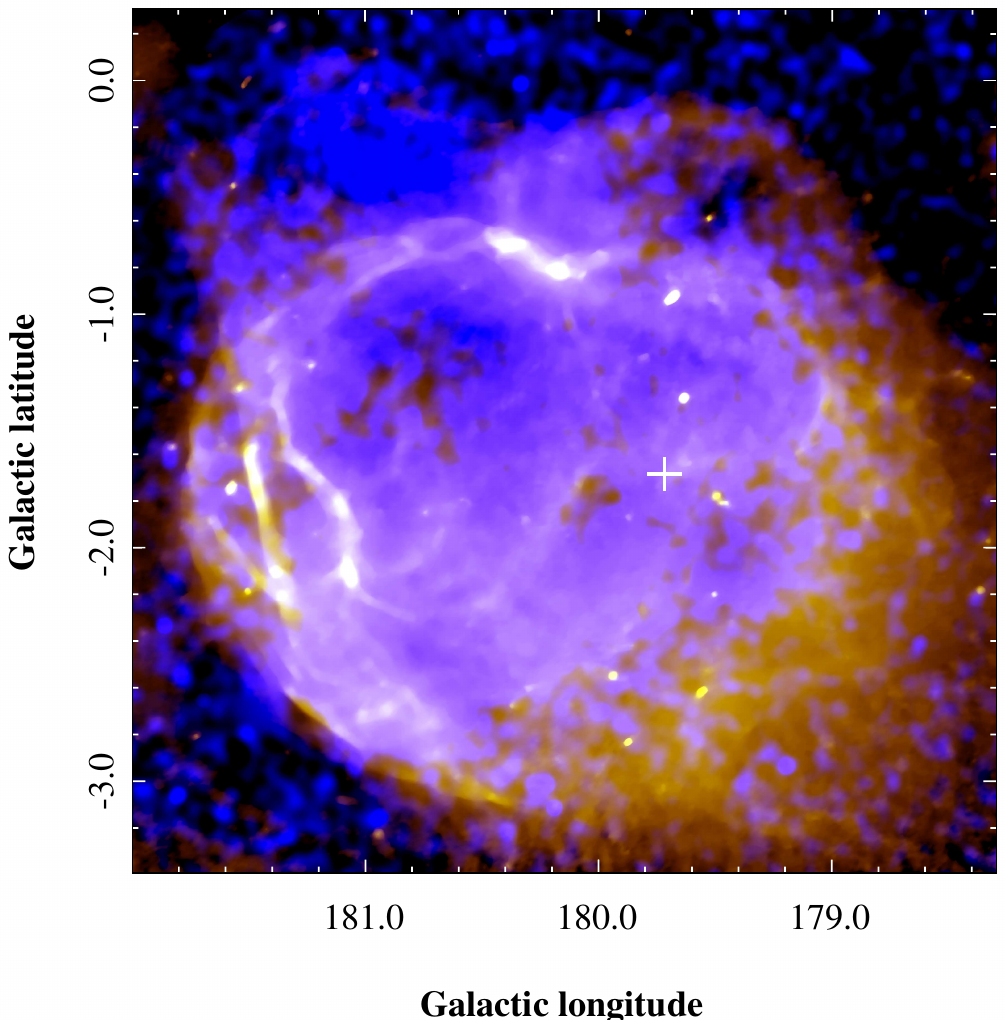}
    \caption{Multiwavelength view of S147 (in Galactic coordinates). \textit{Top} - combined map of the radio (CGPS data at 1.4GHz, red), H$\alpha$ (IGAPS data, green) and broad-band X-ray (0.5-1.1 keV, \srg/eROSITA data) emission. The white cross marks the position of PSR~J0538+2817. \textit{Bottom left} - intensity-saturated X-ray image is shown in blue on top of the wavelet-decomposed H$\alpha$ images, demonstrating that X-ray emission is confined by the H$\alpha$-emitting shell. \textit{Bottom right} - same as the bottom left but with the 1.4 GHz radio emission as a background.}
    \label{fig:multirgb}
\end{figure}

Morphology of the X-ray emission is drastically different from the filamentary and more circular morphology of the H$_\alpha$ and radio emission, as illustrated in Figure \ref{fig:multirgb}(top panel), which combines the same \srg/eROSITA X-ray image (blue) with the H$_\alpha$ image (green) from the IGAPS survey \citep{2021A&A...655A..49G} and radio image at 1.4 GHz (red) from the CGPS survey \citep{2003AJ....125.3145T}. Both latter images were convoluted with a median top-hat filter ($r=40''$) suppressing numerous point sources but leaving diffuse and filamentary emission mostly unaffected.

One can see that X-ray emission is not clearly structured and contains regions of brighter and fainter (by a factor of few compared to the mean one) emission, which are $\sim 1/3-1/10$ of the full size of the nebula ($R\sim100'$) and appear in the form of "blobs" and "depressions" on top of the smooth background emission. No signatures of the global edge-brightening or central peak are visible.

Comparison with the H$_\alpha$ and radio images indicate that some of the bright X-ray regions lie close to the brightest optical and radio filaments (as exemplified by the bright region and filaments North to the nebula's center), while other X-ray bright regions lack prominent counterparts (as the bright X-ray region to the South of the center). On the other hand, the largest cavity in X-ray emission also coincides with the region of fainter optical and radio emission. No signatures of filamentary X-ray emission are visible, and the X-ray emission appears to be well confined by the optical and radio boundaries.

The latter point is even more clearly illustrated by images in the bottom panels of Figure \ref{fig:multirgb}, which show intensity-saturated broad-band X-ray image (blue) on top of wavelet decomposed H$_\alpha$ (left panel, highlighting the filamentary optical emission) and radio (right panel) emission. One can see that X-ray emission reaches the boundaries of the optical and radio emission in the East part of the nebula, while it ends slightly short of it in the West side (appearing more diffuse in the radio and structured on smaller scales in the optical bands). Comparison with the dust and interstellar absorption maps for this region shows no correlations with X-ray morphology indicating that the observed variations are intrinsic to the source itself \citep[cf. also a dedicated discussion in][]{Miltos2024}.

\subsection{Narrow-band imaging}

\begin{figure}
    \centering
    \includegraphics[angle=0,bb=60 175 540 650,width=0.47\columnwidth]{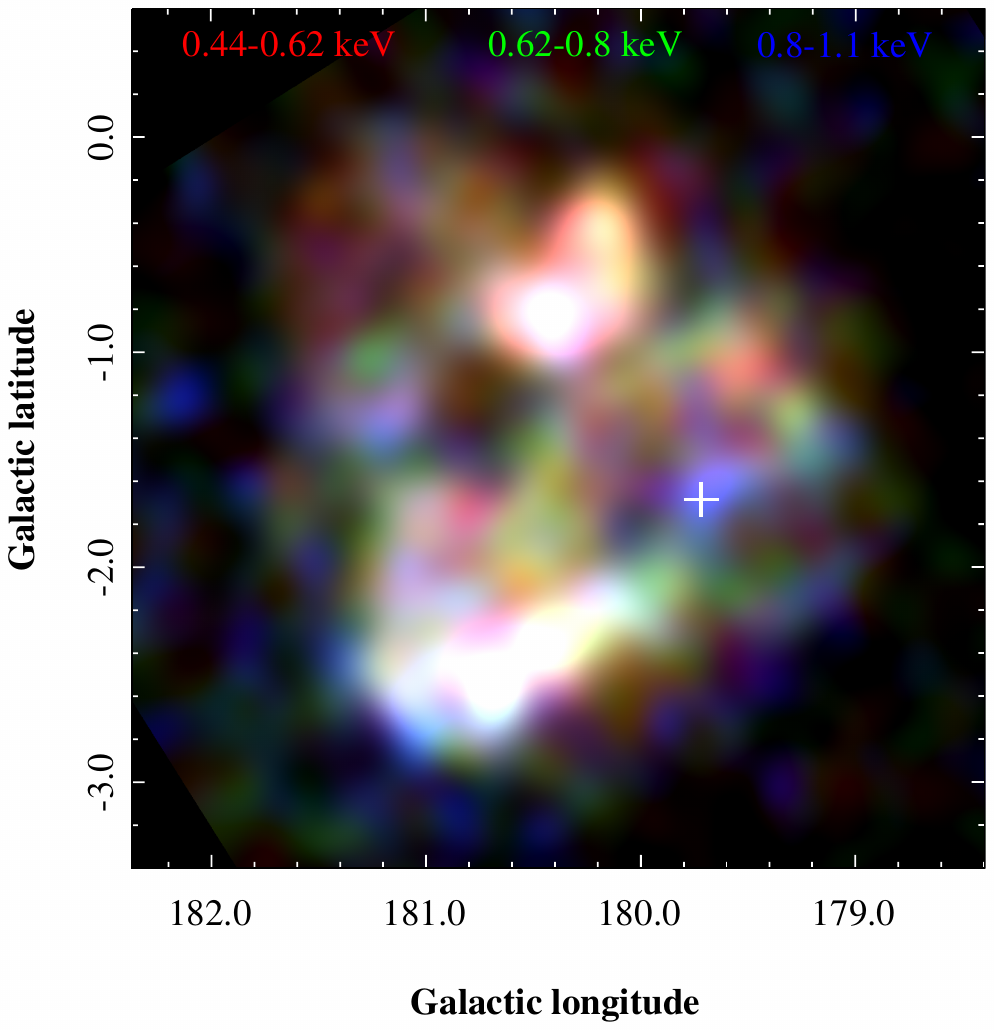}
    \includegraphics[angle=0,bb=60 175 540 650,width=0.47\columnwidth]{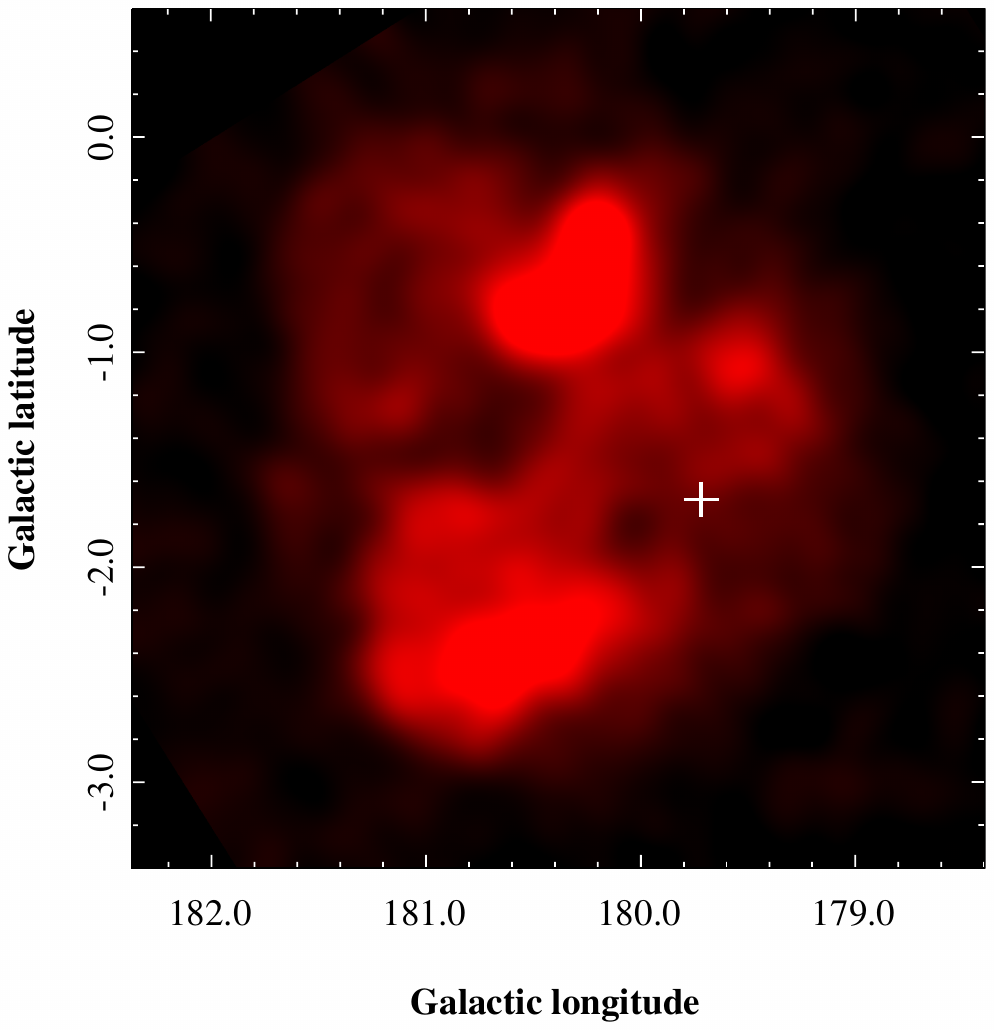}
    \includegraphics[angle=0,bb=60 170 540 645,width=0.47\columnwidth]{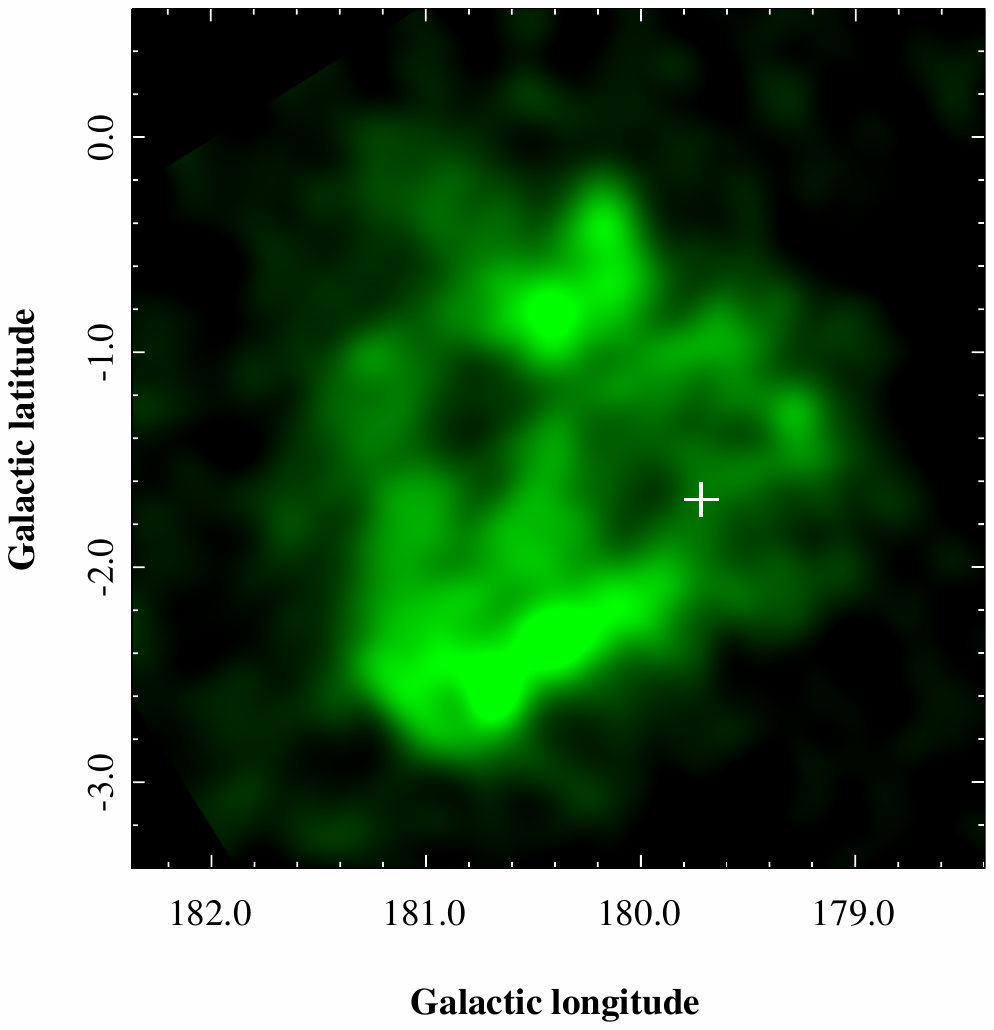}
    \includegraphics[angle=0,bb=60 170 540 646,width=0.47\columnwidth]{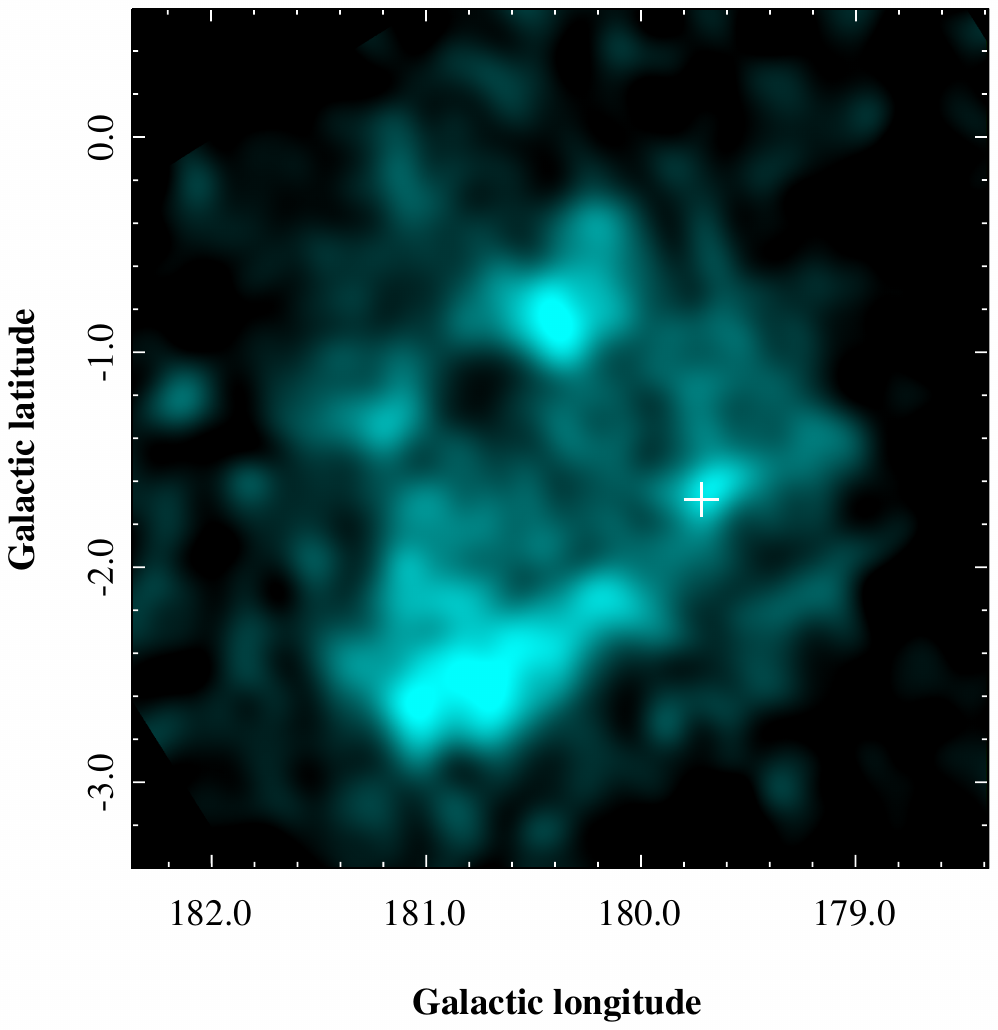}
    \caption{RGB-composite (top left) and individual narrow-band X-ray images covering the 0.44-1.1 keV band. The red, green and cyan (instead of blue for better visibility) images correspond to 0.44-0.62, 0.62-0.8, and 0.8-1.1 keV bands, respectively. These bands encompass the three brightest X-ray emission lines in the spectrum of S147: O~VII, O~VIII, and Ne~IX. }
    \label{fig:xrayrgb}
\end{figure}

Given the inhomogeneous appearance of the X-ray emission from S147 and its complex connection to the emission at other wavelengths, we check for possible correlated spatial variations in its spectral shape by comparing maps accumulated in narrow energy bands centered on the brightest emission lines in the thermal plasma at $kT=0.1-0.3$ keV: O~VII, O~VIII, and Ne~IX. Namely, Figure \ref{fig:xrayrgb} shows images in the 0.44-0.62, 0.62-0.8, and 0.8-1.1 keV bands, as well as their RGB combination.

Although moderate "color" variations are clearly visible on the RGB image, all three images share rather similar morphology.  Emission in the O~VII appears most diffuse, while O~VIII appears more filamentary-structured, and Ne~IX appears somewhat edge-brightened. These observations justify the consideration of the single spectrum of the full nebula as a proxy for the physical conditions of the X-ray-emitting material. Analysis of the spectra extracted from the individual regions is presented in \cite{Miltos2024} and confirms the validity of this assumption.

\subsection{X-ray spectra}

\begin{figure}
\centering
\includegraphics[angle=0,trim=1cm 5cm 0cm 3cm,width=0.99\columnwidth]{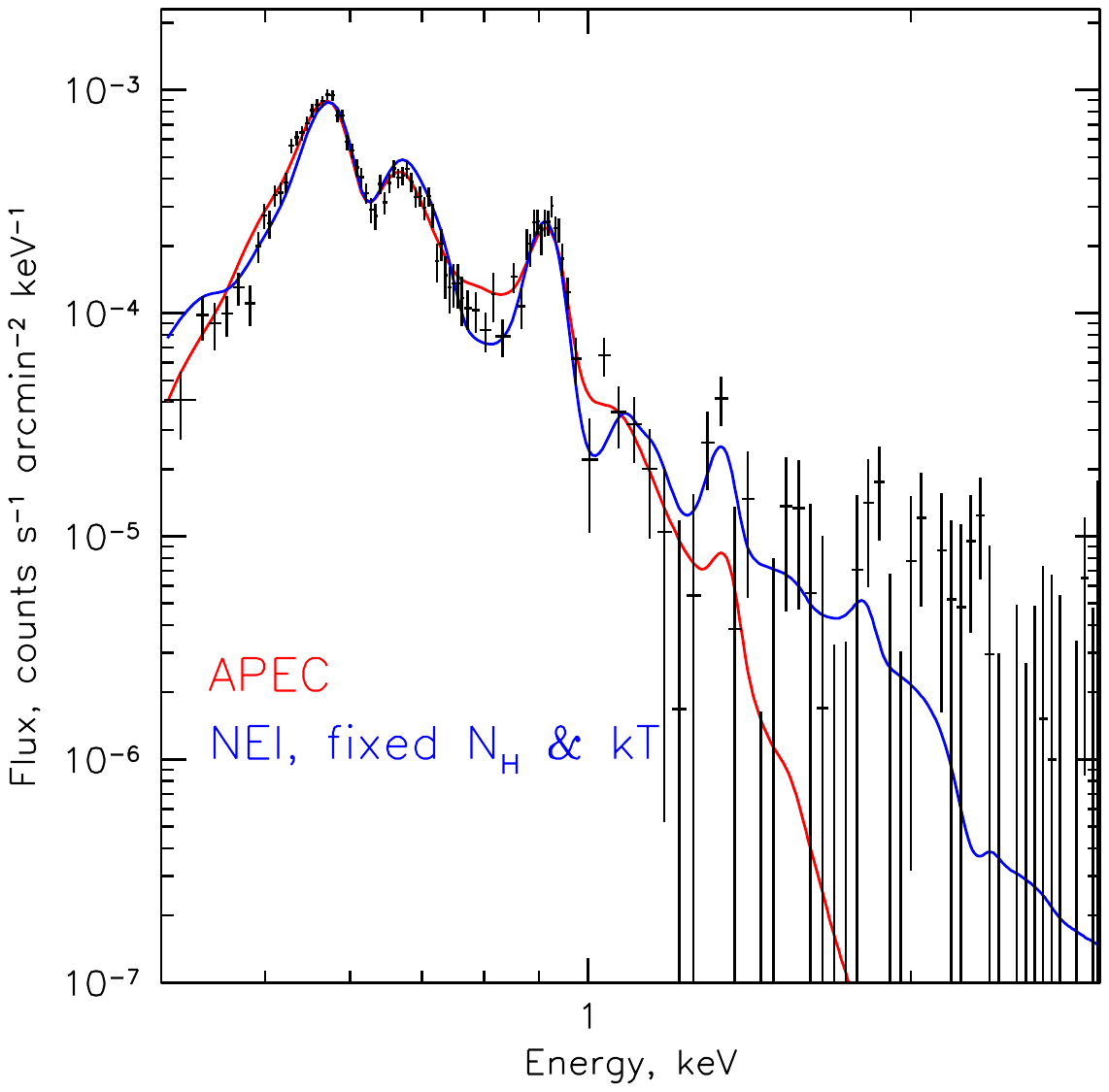}
\caption{X-ray spectrum of the entire S147 SNR. The red line shows the best-fitting \texttt{APEC} model (single-temperature, collisional ionization equilibrium, solar abundance of metals). This model requires a very large absorbing column density, overpredicts the flux near 0.8 keV (where a significant contribution of Fe~XVII line at 826~eV is expected), and underpredicts the Mg~XI line flux unless its abundance (relative to oxygen) is very high.  For comparison, the blue line shows the NEI model with parameters fixed at physically motivated values. Although formally the value of $\chi^2$ is higher than for the APEC model, the lower value of $N_{\rm H}$ and the ability of the model to better describe regions near Fe~XVII and Mg~XI lines make this model an appealing interpretation of the S147 spectrum (see text for details).} 
\label{fig:spec}
\end{figure}

In this section, we consider the most basic characteristics of the entire SNR's X-ray spectrum. 
 The spectrum shown in Fig.~\ref{fig:spec} features three very prominent lines (O~VII, O~VIII, Ne~IX) below 1~keV and weaker lines at higher energies, among which the line of Mg~XI is the most significant.   
 
A starting minimalist's assumption is that for an old SNR, the X-ray emission might be well described by the thermal emission of plasma in collisional ionization equilibrium (CEI). To this end, an absorbed \texttt{APEC} model with a fixed (= Solar) abundance of heavy elements can be used. However, this model fails in two aspects: it requires large absorbing column density ($N_{\rm H}\sim 6\times 10^{22}\,{\rm cm^{-2}}$ that is significantly larger than the expected value  $(0.2-0.4)\times 10^{22}\,{\rm cm^{-2}}$), overpredicts the flux near $\sim 0.8$~keV, and underpredicts the flux in the line of Mg~XI. This is also reflected in the value of $\chi^2$ (366 for 318 d.o.f., see Table~\ref{tab:spec}). In practice, a large value of $\chi^2$ is anticipated given the complexity of the remnant. The integrated spectrum might include several spectral components with different temperatures and metallicities. We, therefore, will not consider the value of $\chi^2$ as a robust quantitative way of ranking complicated models. Instead, we want to demonstrate that for a set of physically motivated parameters, the model can qualitatively reproduce the observed spectrum.  For the S147 model outlined in the Abstract, another spectral model is better motivated. Due to the large size and low density of the gas in the cavity,  the ejecta reheated by the reverse shock some 15~kyr ago can remain hot for a long time and may deviate from CIE. This can happen even if ejecta are mixed with a moderate amount of gas that was present in the cavity. However, a single \texttt{NEI} model converges to a solution that is not very far from the \texttt{APEC} parameters (large $N_H$ and low $kT$), although with the better  $\chi^2$. This is clearly driven by the requirement to describe the O~VII/O~VIII ratio, where statistics is the highest, and given less weight to other parts of the spectrum. One can further try to push ahead NEI scenario by fixing absorbing column density to the expected value of $N_{\rm H}\sim 2.5\times 10^{21}\,{\rm cm^{-2}}$ and the electron temperature to sufficiently high value, e.g. $kT=1\,{\rm keV}$. This model has a higher $\chi^2$ value, but it better describes the spectrum near  $\sim 0.8$~keV and boosts the flux in the Mg~XI line, making this model an appealing interpretation of the S147 spectrum.

\begin{table*}
\caption{The simplest spectral model fits to the spectrum of the entire SNR in the 0.4-3~keV band. The abundance is fixed to solar. Note that for the \texttt{NEI} model with free $N_{\rm H}$ and $kT$, all parameters are highly correlated leading to large uncertainties in their values.
}
\vspace{0.1cm}
\begin{center}
\begin{tabular}{lrrrrr}
\hline
Model & $kT$ & $N_{\rm H}$ & $\tau$ & Normalization & $\chi^2$ (d.o.f.)  \\ 
& keV & $\rm cm^{-2}$ & ${\rm s\,cm^{-3}}$ &{$\rm cm^{-6}\, pc $} & \\
\hline
\hline
\texttt{APEC} & $0.12\pm 0.004$ &$(0.61\pm0.03)\times 10^{22}$& - & $0.95\pm0.28$ & 366 (318) \\
\texttt{NEI} &  $0.14\pm 0.02$ &$(0.57\pm0.07)\times 10^{22}$& $(2.1\pm 0.1)\times 10^{11}$ & $0.33\pm0.28$ & 331 (317)\\
\texttt{NEI}, fixed $kT, N_{\rm H}$ & $1$ & $0.25\times 10^{22}$& $(2.73\pm 0.15)\times 10^9$ &$(0.31\pm0.07)\times 10^{-2}$ & 405 (319) \\
\hline
\hline
\end{tabular}
\end{center}
\label{tab:spec}
\end{table*}

The above analysis was done assuming that the abundance of heavy metals is solar. This assumption can be far from reality if X-ray emission is coming from the ejecta. Since at relatively low temperatures, metals' contribution dominates the X-ray emission, the absolute abundance measurements are difficult, while the relative abundances are more robust. Since the lines of oxygen dominate in the observed spectrum, it is convenient to express abundances relative to oxygen rather than hydrogen. In the models described below (\S\ref{sec:ejecta}) the ejecta abundances of C, N, and Fe (relative to O) are $\sim 30$\% of the Solar values, while the abundances of Ne and Mg vary between between 0.6 and 1.7 (relative to O). Letting the abundance of elements free would make the model much more flexible but less constrained. Since the abundance variations relative to oxygen are not very extreme, we freeze the ratios at the solar values but keep in mind that the line ratios might not be accurately predicted by the model. Another important result of the increased abundance is the difference in the derived emission measure that scales approximately linearly with the abundance (as long as metals dominate the spectrum).

\section{The model: supernova explosion in a wind-blown bubble}
\label{sec:model}

The idea that the supernova might explode in a wind-blown cavity  was first proposed 
for the supernova remnant N132D in LMC \citep{Hughes_1987}.
In that particular case, the cavity was invoked to resolve the disparity between 
the large Sedov expansion age and the much smaller age inferred from optical oxygen-rich filaments.
In the case of S147, we face a somewhat similar age puzzle.

\subsection{S147 age dilemma}

The pulsar J0538+2817 and SNR S147 highly likely have a common origin, which is  
 suggested by    
 the pulsar motion directed from the shell center \citep{Romani_2003}. 
The pulsar distance inferred from VLBI parallax is 1.3 kpc and the 
  proper motion $\mu = 57.9$ mas\,yr$^{-1}$ \citep{Chatterjee_2009}. 
The pulsar offset of $0^{\circ}.605$ combined with the proper motion implies 
 the pulsar kinematic age $t_{\rm kin} = 37.6$ kyr.
With the SNR angular radius of \ha filamentary shell of $100'$ the physical
 shell radius is $r = 39$ pc.

The S147 expansion velocity implied by H$\alpha$ spectroscopic observations is about 100\kms\ \kms\citep{Lozinskaya_1976}. 
This value is consistent with the LAMOST optical spectroscopic survey in the field of S147, which shows 
a symmetric velocity distribution of the line-emitting gas in the range of $\pm 100$ km\,s$^{-1}$ \citep{Ren_2018}. 
Assuming the SNR expansion in a homogeneous medium with a typical density one can derive the age lower limit by applying 
the Sedov solution for the point explosion. 
The inferred SNR age is then 
  $t_{\rm exp} = 0.4r/v = 150$ kyr, which is four times larger than the 
  kinematic age (37.6 kyr). Note, that the radiative expansion regime would produce an even larger age.
 Alternatively, adopting the expansion age of 37.6 kyr and radius of 39 pc one expects 
the shell expansion velocity of $v = 0.4r/t = 407$\kms\ that is four times the observed expansion velocity of 100\kms. 

The disparity between the kinematic and Sedov expansion ages has been recognized by  \cite{Reich_2003} and \cite{Romani_2003}. 
To resolve the age dilemma \cite{Reich_2003} proposed that S147 progenitor exploded 
in a wind-blown bubble (WBB) with the subsequent SNR deceleration after the collision 
with a boundary of the massive swept-up shell.
We consider this scenario as a likely possibility and explore it in more detail.

\subsection{The wind-blown bubble}

As an illustration of the WBB conjecture for the S147, we consider 
the case of a massive SN~II progenitor of about 20\msun. This case is 
relevant because it falls in the range of 9-25\msun\ responsible for neutron stars production \citep{Woosley_2002,Heger_2003}; besides, the 20\msun\ 
mass is high enough to produce an extended WBB for the typical ISM density. 

Commonly, the WBB radius in a homogeneous ISM is estimated using the well-known analytic solution  \citep{Weaver_1977}.  
However, to find a more adequate estimate of the expected radius  one needs to take into account the ISM pressure that is omitted 
in the referred analytic solution. 
We use a thin shell approximation to solve numerically the equations of mass, momentum, and energy conservation of the \cite{Weaver_1977} model with the inclusion of the ISM pressure.
The latter is composed of the thermal pressure of the  Warm Ionized Medium (WIM) component with   
$nT \approx 3000$ cm$^{-3}$\,K \citep{Cox_2005} and relativistic pressure of the interstellar magnetic field $B \approx 3\times10^{-6}$ ($B^2/8\pi \approx 3.6\times10^{-13}$\,dyn\,cm$^{-2}$) combined with the comparable pressure of cosmic rays. 
All in all, one expects the total medium pressure of $\approx10^{-12}$\,dyn\,cm$^{-2}$. 

\begin{figure}
	\includegraphics[width=\columnwidth, trim=3cm 9cm 1cm 7cm]{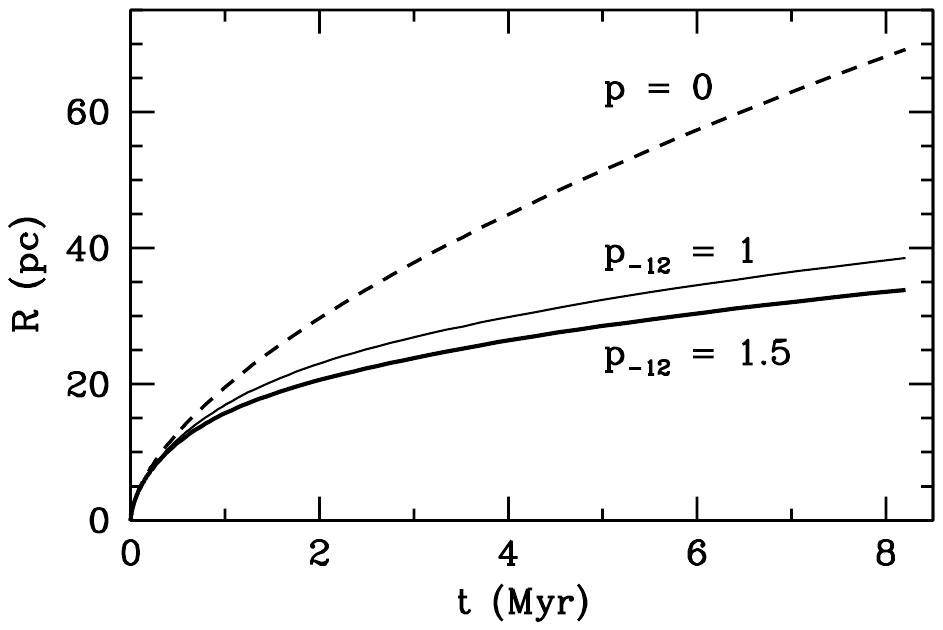}
	\caption{Radius of the wind-driven bubble formed by the 20\msun\ star wind at the main sequence.
	The lines are labeled by the ISM pressure. The model with zero pressure 
	shows the bubble radius based on the analytic solution \citep{Weaver_1977}.
		}
	\label{fig:bubev}
	\end{figure}

Major properties of the model bubble for the ISM density of 0.3\cmq\ and three choices of the ISM pressure are illustrated by Figure \ref{fig:bubev} and Table \ref{tab:bubmod}. The Table contains input parameters: stellar mass, main-sequence lifetime \citep{Schaller_1992}, mass loss rate, wind velocity, ISM pressure and
characteristics at the end of the main sequence: the bubble radius, bubble density, and the shell swept-up mass.
The adopted wind parameters correspond to the main sequence O-star star with the mass 
of 20-25\msun\ \citep{Howarth_1989}.
The main sequence wind (0.7\msun) is thermalized in the termination shock 
with the radius $r_t \approx 3$\,pc and uniformly fills throughout the cavity volume.

At the He-burning stage (0.79 Myr) the progenitor becomes red supergiant (RSG) and 
loses matter via the slow wind ($\sim 15$\kms) with the mass loss rate of about $10^{-6}\,{\rm Myr}$. Almost all the RSG wind  ($\approx 1$\msun) is expected to be swept up,  due to the bubble counter-pressure, into the dense shell with a radius of 
$\approx 1$\,pc, significantly smaller than the bubble radius.

\subsection{Effect of cloudy ISM}

The model of the almost empty WBB in the homogeneous medium is an idealization. 
Indeed, we see a signature of possible interaction of S147 with a 
large scale ($\sim 40$\,pc) cloud that is hinted by the brightening in \ha, radio, and gamma-rays towards the southern part of the SNR shell at around coordinates $\alpha = 5^h42^m,  \delta = 26^o35'$.
The H\,I 21~cm data \citep{Dickey_1989} imply that the mass spectrum of interstellar 
clouds, both molecular and diffuse atomic, is $dN/dm \propto m^{-\gamma}$ with $\gamma \sim 2$. 
One expects therefore to find among $\sim 2\times10^3$\msun\ of ISM swept up by S147 a significant number of parsec size clouds with the density of 10-100\,cm$^{-3}$.

The key question is whether some of these clouds avoid disruption by the dense bubble shell and end up inside the bubble.
A cloud, in order to cross the shell safely, should have the column density $N_c \approx n_cr_c$ significantly larger than the shell 
column density $N_s = nr_b/3 \sim 10^{19}$\,cm$^{-2}$ (for $n = 0.3$\,cm$^{-3}$ and $r_b = 30$\,pc). 
A typical cloud that meets this requirement
might have the radius of $\sim 1$\,pc and density of $n_c = 30$\,cm$^{-3}$, in which case $N_c \approx 10^{20}$\,cm$^{-2}$, by a factor of 10 greater than  $N_s$.
Typical column densities of molecular clouds are 1--2 orders of magnitude higher \citep[e.g.,][]{Larson1981,Heyer2015}.
One, therefore, expects to find a significant amount of parsec size interstellar clouds engulfed by the bubble. 
While these clouds do not affect the bubble dynamics,  
expanding supernova ejecta  will crush clouds, and disperse 
them into fragments with eventual mixing, thus producing overdensities after their thermalization by the reverse shock. 

A cloud with the radius of 1\,pc and H density of 30\,cm$^{-3}$ has the mass 
of $\approx 3$\msun.
 Whether such clouds survive a complete stripping due to photoevaporation 
 before the bubble shell crossing? 
 The cloud photoevaporation depends on the cloud mass $m$ and ionizing flux $Q/R^2$ as $\dot{m} \propto m^{3/5}(Q/R^2)^{1/5}$ 
  \citep{Bertoldi_1990}. Using the relevant expression from this paper we 
 find that 
the cloud with the mass $> 1$\msun\ at the distance $r \gtrsim 20$\,pc 
survives the photoevaporation during the main sequence lifetime of 20\msun\ star and thus has a chance to find itself in the bubble.

\begin{table}
	\caption[]{Parameters of bubble model.\\
	 The upper part shows adopted input parameters. Lower part 
 displays bubble radius, density, and swept-up mass for the adopted 
 ISM pressure.}
	\begin{tabular}{l|c|c|c}
  \hline
  Physical quantity  & \multicolumn{3}{c}{Numerical value} \\
		\hline
		$M$\quad (\msun)                 &   \multicolumn{3}{c}{20} \\ 
	$t_{ms}$\quad (Myr)	       &  \multicolumn{3}{c}{8.1}\\       
	 $\dot{M}$\quad (\msyr)   &   \multicolumn{3}{c}{$10^{-7}$} \\
	 $v_w$\quad (\kms)       &  \multicolumn{3}{c}{1500}\\
  \hline
     $p$\quad ($10^{-12}$\,dyn\,cm$^{-2}$)   &\qquad   0  & 1  & 1.5 \\      
  \hline     
   $r_b$\quad (pc)            &\qquad   69        &   38.5   &  33.9 \\
   $\rho_b$\quad($10^{-28}$\gcmq) 	 &\qquad 0.4 &  2.3  &  3.4\\
   $M_{ds}$\quad($10^3$\msun)     &\qquad 14.6   & 2.5 & 1.7 \\
 \hline
	\end{tabular}
	\label{tab:bubmod}
\end{table} 

\subsection{Ejecta}
\label{sec:ejecta}

\begin{figure}
\centering
\includegraphics[angle=0,trim=1cm 5cm 0cm 3cm,width=0.99\columnwidth]{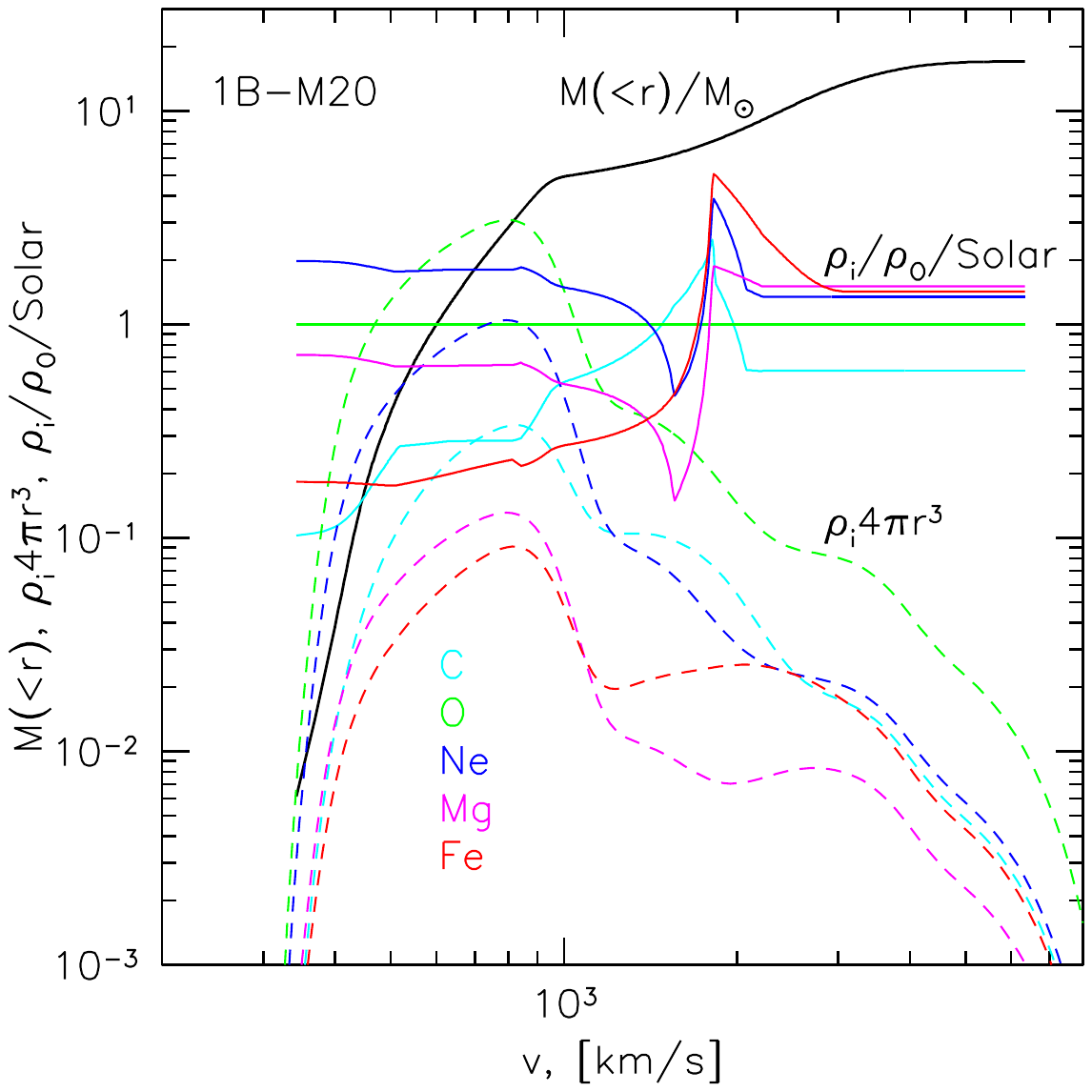}
\caption{Ejecta structure for the 1B-M20 model. The black solid line shows the enclosed ejecta mass (for a given velocity). The colored dashed lines show what velocities make the largest contribution to the mass of a given element, namely, the quantity $4\pi r^3\rho_i$, where $\rho_i$ is the mass density of the $i$-th element, color-coded as shown in the legend. The colored solid lines show the mass density of the $i$-th element relative to the mass density of oxygen, normalized by the Solar value. }  
\label{fig:ejecta}
\end{figure}

To follow the development of the SN explosion within the extended WBB, we
   explode the pre-SN model based on a 20\msun\ progenitor evolved by
   \citet{Woosley_2002} from the main sequence up to the onset of core collapse.
Unfortunately, \citet{Woosley_2002} used so high mass-loss rate that the
   resultant mass of the corresponding pre-SN was 14.7\msun.
In turn, we use a moderate mass-loss rate and assume that the 20\msun\ progenitor
   lost $\approx$0.7\msun\ at the main-sequence stage and $\sim$1\msun\ during
   the RSG phase.
So for our problem, we take the pre-SN model of 18.5\msun.
To construct the relevant pre-SN model, we modify the original pre-SN
   model of 14.7\msun\ by increasing the mass of the hydrogen-rich envelope
   up to 12.4\msun, preserving both the helium core of 6.1\msun\ and the shape
   of the profile of density in the hydrogen-rich envelope except for the
   interface between the helium core and the envelope.
The obtained model is exploded with the energy of $10^{51}$\,erg by a piston
   at the outer edge of the central collapsing core of 1.46\msun. 
The artificial mixing applied to the pre-SN model mimics the intense 3D turbulent
   mixing at the (C+O)/He and He/H composition interfaces occurring during the
   explosion \citep{Utrobin_17}.
In addition, radioactive $^{56}$Ni with mass of 0.07\msun, typical for type II
   SNe, is mixed artificially in velocity space up to nearly 3000\kms.
Figure~\ref{fig:ejecta} shows the profiles of important chemical elements
   in the freely expanding ejecta.
In light of an analysis of the content of these elements, we investigate
   the pre-SN models for the progenitor masses in the range from 18\msun\
   to 22\msun\ \citep{Woosley_2002}.
In particular, we find that the total Mg mass in the ejecta for the 19\msun\
   progenitor is twice as large compared to that of the 20\msun\ progenitor.

\begin{figure}
\centering
\includegraphics[angle=0,trim=1cm 5cm 0cm 3cm,width=0.99\columnwidth]{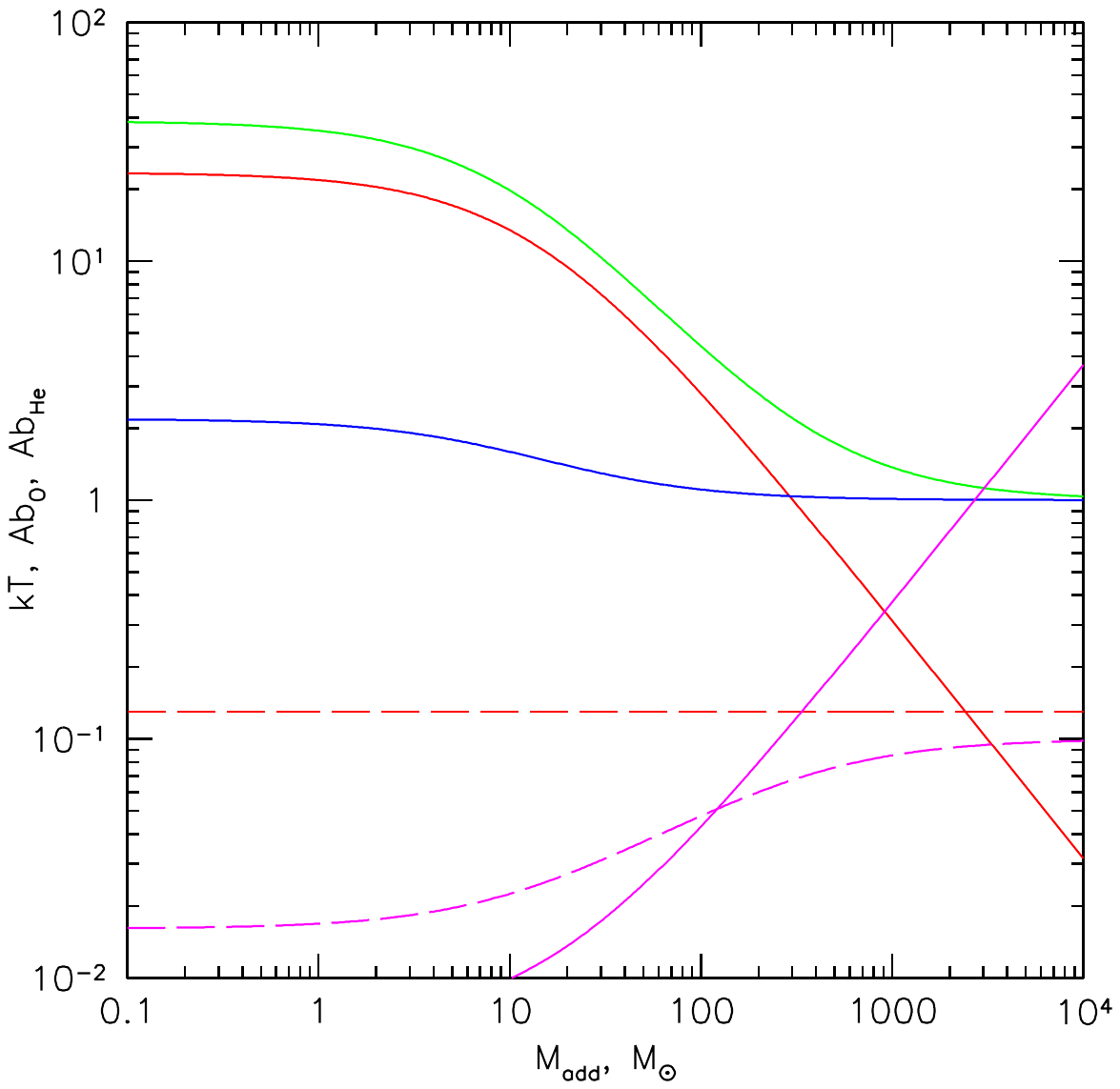}
\caption{Dependence of the mean temperature (the solid red line) and He (blue) and O (green) abundances when ejecta are mixed with a given mass ($M_{\rm add}$) of the ISM. For the temperature calculations, we assume that the initial energy $E=10^{51}\,{\rm erg}$ goes entirely into the gas heating, which is fully ionized. For the abundance calculation, we assume solar abundances for the ISM and the 1B-M20 model for the ejecta. For comparison, the horizontal dashed line is the best-fitting temperature for a one-temperature APEC model. The intersection of the red dashed and solid lines suggests a large added mass $\sim 2-3\times 10^3\,M_\odot$. Similarly, the dashed magenta line shows the density derived from the normalization of the same model, taking into account the variable abundance of elements (focusing on O). The intersection of the magenta dashed and solid lines suggests a much smaller added mass $\sim 10^2\,M_\odot$. This discrepancy clearly demonstrates that the above set of assumptions (CEI, APEC spectral model, complete mixing of the ejecta and the gas inside the cavity,  and $T_e=T_i$) is likely violated.}
\label{fig:madd}
\end{figure}

\begin{figure}
\centering
\includegraphics[angle=0,trim=1cm 5cm 0cm 2cm,width=0.99\columnwidth]{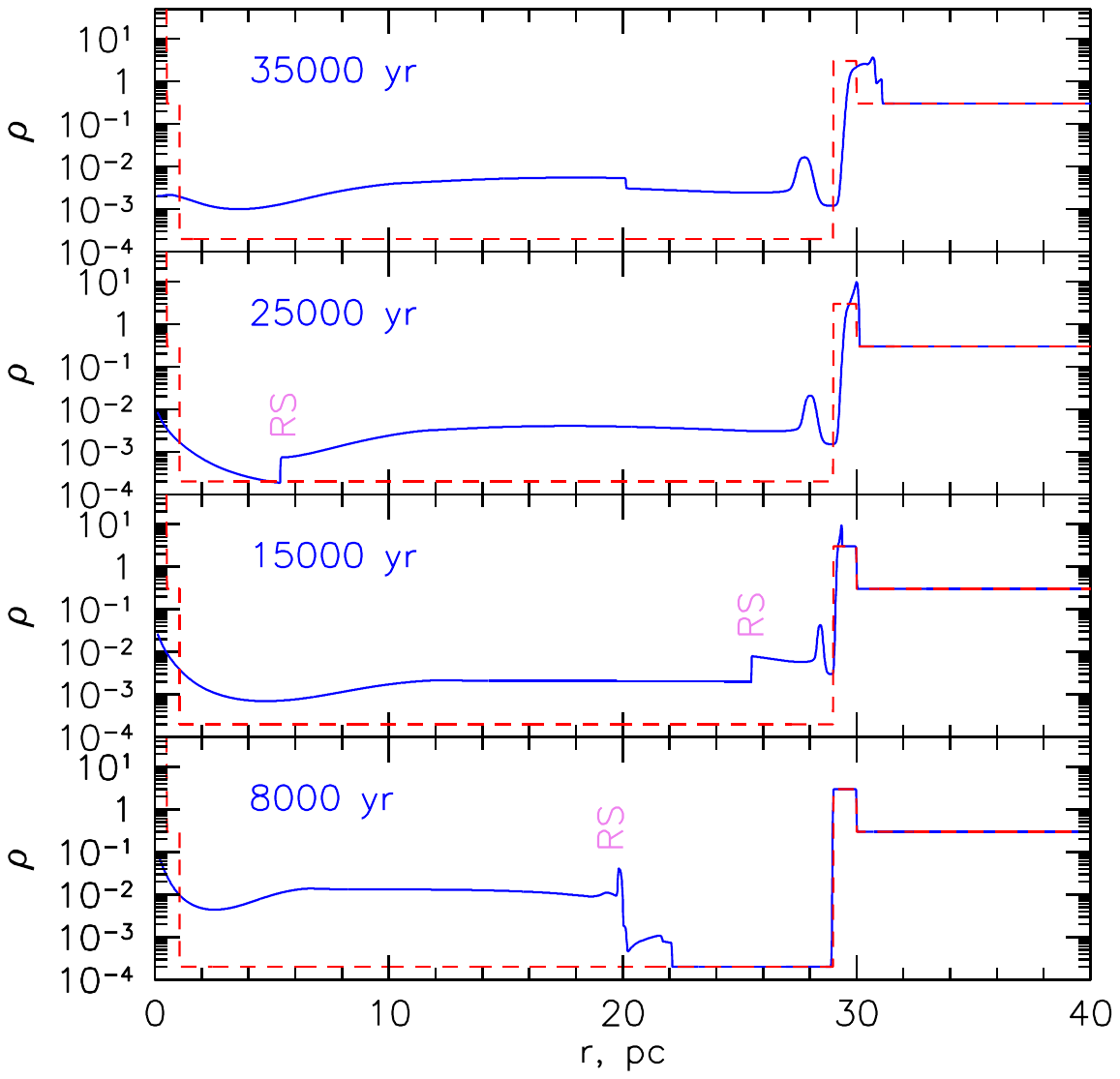}
\caption{Propagation of the shock through a low-density cavity bound by a high-density shell.The density is in units of $m_p\,{\rm cm^{-3}}$. The dashed line shows the initial density profile, while the blue line shows the density profile evolution. When the forward shock moves through the low-density gas the reverse shock moves outwards. After the collision with the dense shell, the reverse shock propagates deep into the ejecta. These simulations are non-radiative, which is a reasonable approximation for the ejecta and the cavity. The forward shock in the dense gas is radiative, i.e. its structure is not correctly reproduced by these simulations. A jump at 20~pc in the upper panel is the reflected reverse shock that appears in a spherically symmetric 1D model. 
}
\label{fig:hyd}
\end{figure}

\begin{figure}
\centering
\includegraphics[angle=0,trim=1cm 5cm 0cm 2cm,width=0.99\columnwidth]{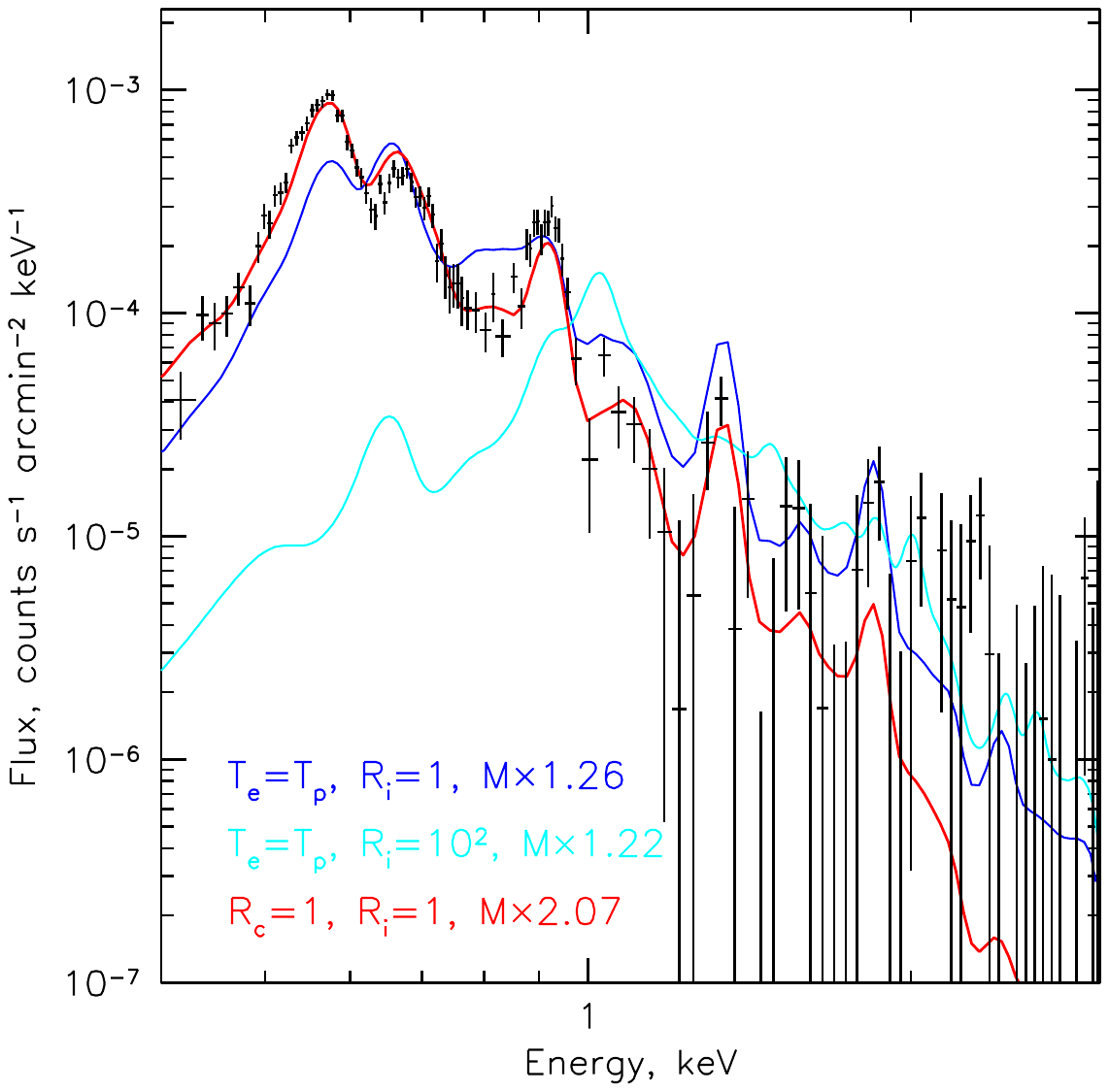}
\caption{SNR spectra based on the 1D hydro model shown in Fig.~\ref{fig:hyd} at $t=3\times 10^4\,{\rm yr}$. The black crosses are the same data points as in Fig.~\ref{fig:spec}. 
The models (blue, cyan, and red curves) differ in the level of deviation from the CIE state with $T_e=T_i$.  In particular, the blue curve corresponds to instantaneous electron/proton equilibration, i.e. $T_{\rm e}=T_{\rm p}=T_{\rm i}$ and self-consistently calculated evolution of the ionization balance. The cyan curve shows the case when the rate of all collisional processes is artificially increased by a factor of 100 (i.e. this case is close to the CIE state with $T_e=T_i$. The red curve shows the case when most of the energy downstream of the shock goes into protons and the electron/proton equilibration proceeds via pure Coulomb collisions.   In all cases, the ionization balance is followed in each shell according to the time evolution of electron temperature. The model spectra were integrated along the line of sight at a projected distance of $0.35$ times the radius of the SNR. }  
\label{fig:hyd_spec}
\end{figure}

\begin{figure}
\centering
\includegraphics[angle=0,trim=1cm 7cm 2cm 4cm,width=0.99\columnwidth]{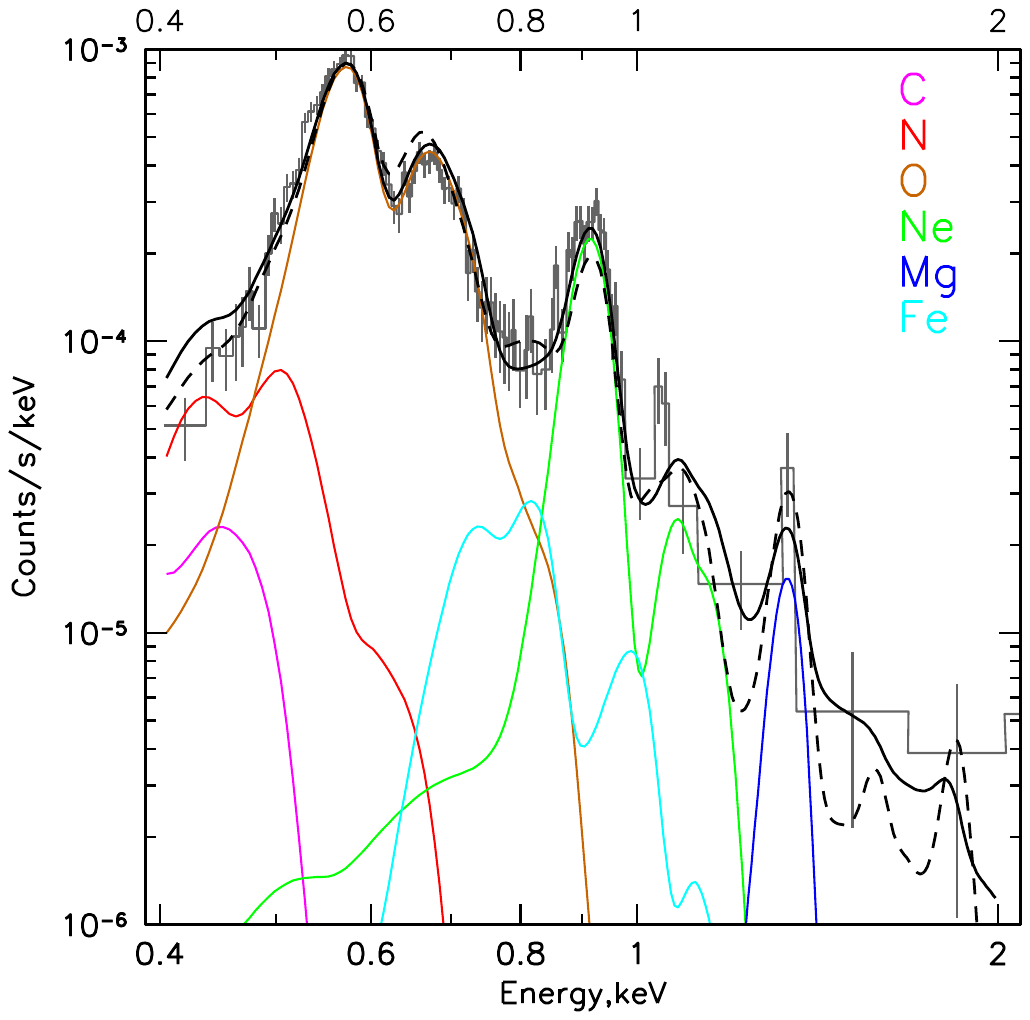}
\caption{Contribution of individual elements to the S147 X-ray spectrum in the \texttt{NEI} model (black solid line). The dashed line shows the spectrum predicted by the 1D-hydro model.}  
\label{fig:spec_lines}
\end{figure}

\begin{figure}
\centering
\includegraphics[angle=0,trim=1cm 5cm 0cm 2cm,width=0.99\columnwidth]{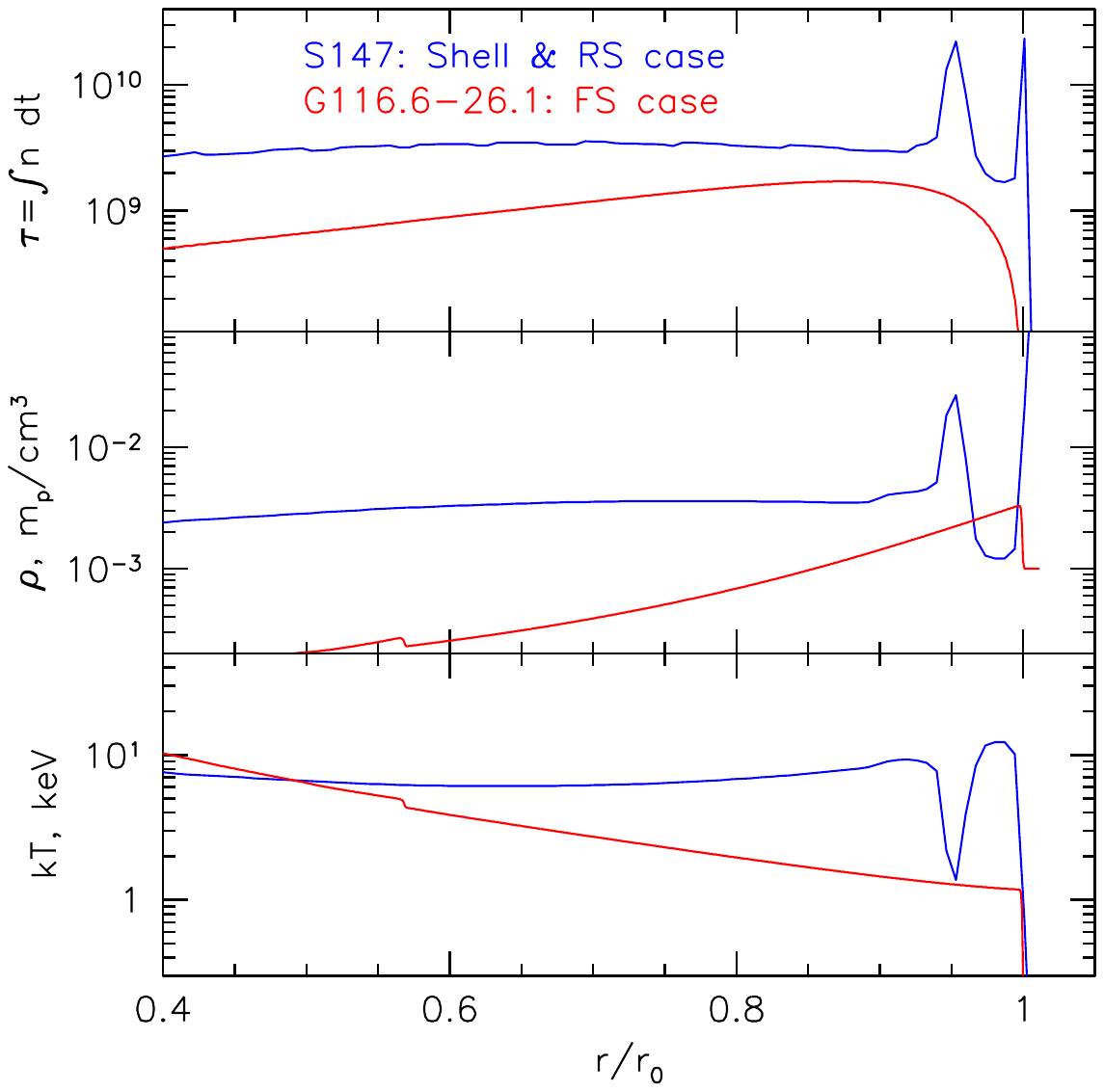}
 \caption{Radial profile of the ionization parameter $\tau=\int n dt$ (the blue line, top panel) based on 1D model, when most of the X-ray emission is coming from the gas reheated by the reverse shock following the collision of the ejects with the dense shell. The radius is normalized by the characteristic radius $r_0$, which is approximately equal to the radius of the forward shock. Small amplitude irregularities in the black curve at $r/r_0\lesssim 0.8$ are the artifact of $\tau$ calculations. For comparison, the red curve shows the case when the X-ray emission is mostly due to the gas heated by the forward shock (the parameters for G116-26.1 are used here). Two spikes in $\tau$ are clearly visible at large radii. The outmost spike is likely an artifact of our non-radiative run near the inner boundary of the dense shell. The other spike is due to the less dense gas (outer layers of ejecta) that has a relatively high temperature and can contribute to X-ray emission. In this region, oxygen might be completely ionized, but the lines of Si and Fe might be present, as illustrated in Fig.~\ref{fig:limb}. The middle and bottom panels compare the model density and temperature profiles for these two SNRs.}  
\label{fig:tau}
\end{figure}

\begin{figure}
\centering
\includegraphics[angle=0,trim=1cm 5cm 0cm 2cm,width=0.99\columnwidth]{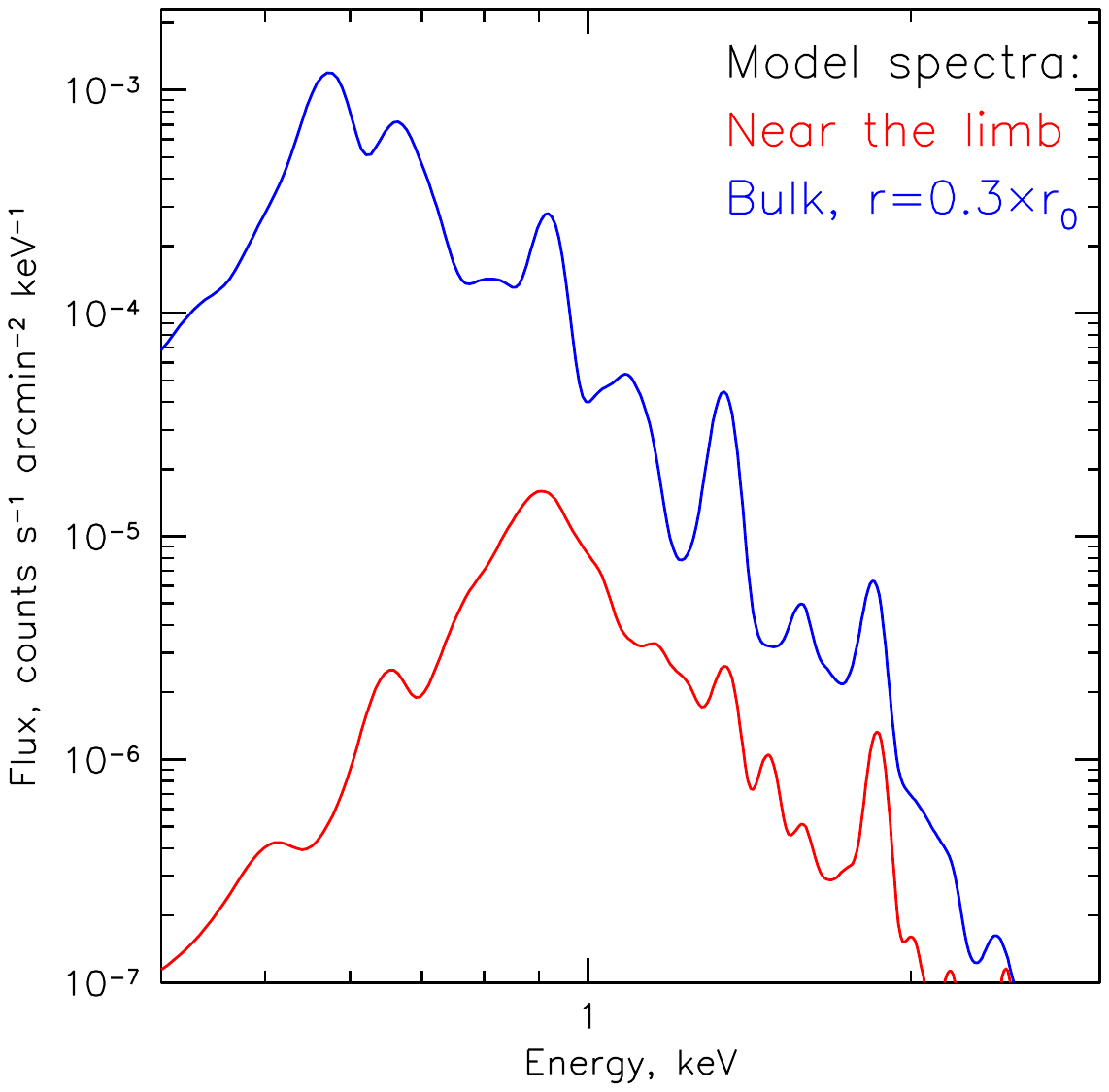} 
\caption{Projected model spectra for the SN-in-the-WBB scenario near the limb (red) and towards the inner parts of the nebula (blue). The gas near the limb has a larger ionization parameter than the bulk of the volume and, accordingly, lacks oxygen lines, but features the lines of Ne, Fe, and Si. In reality, S147 is far from spherical symmetry and a mixture of components with a range of ionization parameters should be observed.}  
\label{fig:limb}
\end{figure}

\subsection{Morphology of the H$\alpha$ emission}

 No doubt that the intricate spaghetti-like structure is a complicated manifestation of the late radiative stage of the SNR liable to different instabilities \citep{1998ApJ...500..342B}.
The forward shock temperature assuming $T_e = T_i$ is $T = 1.36\times10^5v_7^2\quad\mbox{K}$.
The cooling time in the postshock gas for the shock with the speed of 100\kms\ \citep{Lozinskaya_1976} is 
\begin{equation}
t_c = \frac{3kT}{4\Lambda n} = 500v_7^2/n \quad \mbox{yr}\,,
\end{equation}
where $\Lambda(T) \approx 10^{-21}$\,erg\,s$^{-1}$\,cm$^3$ \citep{Sutherland_1993}.
With $t_c << t$ the forward shock is indeed radiative. 

\cite{Pikelner_1954} proposed that the \ha\ filament is a density enhancement at the shock wave intersection that arises in a corrugated shock interacting with an interstellar cloud.
\cite{Kirshner_1979} consider two options for the ``filament'' --- ropelike structure and sheet viewed edge-on --- and 
conclude that filaments are ropelike structures.

We believe that the S147 structure includes both options. 
The limb brightening is apparent at the shell boundary. 
However, ropelike filaments seem to be responsible for the majority of spaghetti structures. 
We share the view of \cite{Pikelner_1954} that ropelike filaments originate from the 
intersection of shock waves. 
The foamy structure of the global radiative shock favors crossing of neighboring 
protrusions.
The shock wave protrusions could originate from either 
instabilities \citep{Blondin_1998} or the shock wave propagation in the essentially 
inhomogeneous ISM.
The latter possibility is illustrated by the foamy structure 
of a model SNR produced by three-dimensional hydrodynamic simulations of a supernova expansion in the inhomogeneous ISM \citep{Martizzi_2015}.

The almost circular shape of some filaments suggests that they are produced by the intersection of a 
convex shock protrusions with a more or less plane shock. 
In this case, one expects comparable radial and tangential \ha\ velocity components that should cause an
azimuthal dependence of the radial velocity along the circular filament with large impact parameter $p \gtrsim 0.5$ (distance from the shell center in units of the shell radius).
The expected behavior of the filament radial velocity could explain the absence of the anti-correlation between the \ha\ radial velocity and the impact parameter that was established earlier \citep{Lozinskaya_1976,Kirshner_1979}.

\subsection{Spectral model motivated by 1D hydro model}
The 1D model shown in  Fig.~\ref{fig:hyd} illustrates the overall concept of the model discussed here. While the parameters used in the model might not match the properties of the S147 progenitor and environment accurately, it is interesting to compare the X-ray spectra expected in this simple model with the data. This can be done in several steps. One needs to calculate the time evolution of the electron temperature in each radial shell, evolve ionization fractions, calculate the energy-dependent emissivity,  and integrate it along the line of sight. To this end, we follow the same approach as in \cite{2021MNRAS.507..971C,2023MNRAS.521.5536K}. Here, for spectra calculations, the \texttt{MEKAL} model \citep{1985A&AS...62..197M,1995ApJ...438L.115L} as implemented in the \texttt{XSPEC} package \citep{1996ASPC..101...17A} was used in combination with the evolving ion fractions. For the electron temperature, the following three cases have been considered.
\begin{enumerate}
\item $T_e(r,t)=T_h(r,t)$, where $r$ and $t$ are the shell radius and time, respectively, $T_h(r,t)$ is the fluid temperature in the hydro run for a reasonable choice of the mean molecular weight $\mu$. This case corresponds to the equal temperature of all particles (the blue line in Fig.\ref{fig:hyd_spec}).
\item In the second case, we assume that most of the energy downstream of the shock goes into protons and electron/proton equilibration proceeds via pure Coulomb collisions (the red line).
\item Finally, the last model also involves Coulomb collisions, but the rate of collisions is artificially increased by a factor of $100$ (the cyan line). This model is expected to be close to the first model.
\item For the 2nd version of the model-spectrum plot: Finally, the last model assumes $T_e(r,t)=T_h(r,t)$ similar to the first model, but the rates of ionization/recombination are artificially increased by a factor of $100$ (the cyan line). This model is expected to be closer to CIE than other models.
\end{enumerate}
Fig.~\ref{fig:hyd_spec} shows that the second model fits the data surprisingly well, given the simplicity of the model. In this figure, the model spectrum corresponds to the projected distance $r_p=0.35 r_{s}$ from the center of the SNR, where $r_s$ is the forward shock radius, although the dependence on the value of $r_p$ is rather weak. For solar abundance of heavy elements, the normalization of the predicted surface brightness is a factor of $\sim 2$ lower than the observed spectrum. Increasing abundance in the model by a factor of 2 brings the model very close to the observed spectrum. Large values of the abundance would overpredict the flux.  As follows from Fig.~\ref{fig:madd}, the full mixing scenario with the small amount of mass (as used in our 1D model) would lead to a larger abundance of Oxygen. This suggests that in S147, most of the X-ray emission is associated with the outer layers of the ejecta. The total energy "problem" apparent in Fig.~\ref{fig:madd} is easily resolved in the model with $T_i\gg T_e$ with the standard rates of temperature equilibration via Coulomb collisions. 

The shell model adopted here has an interesting implication for the temperature and ionization structure of the SNR as seen in X-rays. Indeed, since much of the X-ray emission is expected to come from the gas reheated by the reverse shock, the profile of the ionization parameter $\tau=\int n dt$ can be different from the standard case of a Sedov-like solution, where the forward shock is responsible for X-ray emission. This is illustrated in Fig.~\ref{fig:tau}, which shows the radial profiles of $\tau$ for S147 and G116-26.1. The latter is used to illustrate the forward-shock-dominated case. In both cases, the calculation of $\tau$ for a given fluid element begins when this fluid element goes through a shock and is heated to temperatures $\sim 1\,{\rm keV}$. 

For S147, the value of $\tau$ is almost constant through the entire ejecta, unlike the G116-26.1 case. In addition, S147 might have a spike in $\tau$ behind the position of the forward shock, due to the outer layers of ejecta that were reheated by the second reverse shock (see Fig.~\ref{fig:hyd}. Depending on the initial gas density in the cavity, such a spike might arise due to this gas. What is needed is a moderately dense gas that can be reheated by reverse shock to high temperatures. This combination of parameters would drive $\tau$ up and power X-ray emission.  
If this model is correct, near the limb of S147 one can expect to find hotter and more highly ionized gas than in the rest of the nebula. This is illustrated in Fig.\ref{fig:limb}.

\subsection{Abundances}
Most of the above analysis was done assuming solar abundance (or solar mix of abundances) in the X-ray emitting gas. As illustrated by Figs.\ref{fig:ejecta} and \ref{fig:madd} the abundance of oxygen relative to hydrogen is sensitive to the amount of "cavity gas" added to the ejecta and to the assumption that entire ejecta are mixed with this gas and all this gas contributes to the observed X-ray emission. Given the complexity of S147, this is likely a severe simplification. On the other hand, the variations of abundances relative to oxygen are less extreme, in the range of 0.3-2 for elements from C to Fe, when averaged over the entire ejecta. Fig.~\ref{fig:spec_lines} shows the contributions of individual elements to the \texttt{NEI} model with solar abundances. Clearly, the underabundance of C and N in the adopted ejecta model might affect the absorbing column density derived from the X-ray spectra. For the heavier elements, like Ne, Mg, and Si, a factor of $\sim 2$ uncertainty (or larger) is feasible. Our conclusion here is that allowing abundance variations, coupled with uncertainties in $T_e$ and non-equilibrium ionization would allow for an even better fit to the observed X-ray spectra, although the degeneracy between parameters would lead to large uncertainties in the derived parameters. 

\subsection{Non-thermal components in S147}

While no signature of non-thermal X-rays was found in the eROSITA data analysis it is still 
worth discussing the presence of non-thermal components in multiwavelength data since these 
are needed to justify the SNR model presented above. Indeed, the shocks of velocity $\gtrsim$ 100 \kms at the current stage of S147 evolution are not fast enough to allow particle acceleration above TeV energies with magnetic field amplification which are needed to extend the synchrotron radiation spectra to X-rays as it is likely the case in young SNRs (see e.g. \citet{CR_B18,vink20}). However, GeV regime electrons and protons can be accelerated and confined in S147 providing the observed radio and gamma-ray emission, and may contain a sizable energy density $\sim$ 10\% of the total gas pressure.  

The radio image of S147 is composed of "filaments", associated with H$\alpha$ 
filaments, and a diffuse component. The appearance of radio filaments has the same origin as the H$\alpha$ filaments, i.e., the limb brightening.
The overall integrated flux density is $34.8\pm 4$ Jy at 11 cm \citep{2008A&A...482..783X}.
 The spectrum is flat 
($\alpha = -0.30\pm0.15, F \propto \nu^{\alpha}$) in the range of $\lesssim 1.5$ GHz and gets steeper at higher frequencies \citep{2008A&A...482..783X}.

The observed flux at 11 cm can be described based on a simple model of a  homogeneous spherical radio-emitting shell filled with cosmic rays and magnetic fields.
The outer radius of the SNR is  $R_2 = 39$ pc and the inner radius $R_1 = 0.9R_2$ (adopting the SNR distance of 1.3 kpc). For initial estimations, one can assume the equipartition between cosmic rays and magnetic field energy density ($U_{cr} = B^2/8\pi$) with a typical electron-to-proton ratio $\xi_{e,p} = U_e/U_p = 0.01$. 
The energy spectrum of relativistic electrons responsible for the radio emission is assumed to be $dN/dE = KE^{-p}$ with $p = 2$ in the GeV range consistent with that expected in the diffusive shock acceleration model. 

The observed flux at 11 cm then is reproduced for the total energy of relativistic component $E_{rel} = 9\times10^{49}$ erg, which suggests 
the energy density $U_{rel} \approx 4.5\times10^{-11}$ erg\,cm$^{-3}$ and 
magnetic field of about 24\,$\mu G$.  
 Remarkably, the relativistic pressure in this model $p_{rel} \approx 1.5 \times10^{-11}$ dyn\,cm$^{-2}$ turns out a sizable fraction of the upstream dynamical pressure.
Indeed, for $v_s = 100$ km\,s$^{-1}$ and preshock density 
$n_0 = 0.3$ cm$^{-3}$ 
 the dynamical pressure is $p_{dyn} = \rho_0v^2 \approx 7\times10^{-11}$ dyn cm$^{-2}$ with the ratio $\xi = p_{rel}/p_{dyn} = 0.2$. The simple estimations above assumed a homogeneous radio-emitting shell while the radio images revealed a filamentary structure of the SNR which probably indicates a highly intermittent structure of magnetic fields.

Non-thermal pressure estimated from radio emission of relativistic electrons in SNR can be compared with that derived from the gamma-ray data. Fermi-LAT observations of S147 by \citet{2012ApJ...752..135K} in the energy range 0.2-10 GeV revealed an extended gamma-ray source of luminosity $\sim 1.3\times 10^{34}$ erg~s$^{-1}$ at the assumed distance 1.3 kpc. They pointed out an apparent correlation of the gamma-ray emission with the \ha filaments and found no signal associated with the pulsar PSR J0538+2817. These authors concluded that re-acceleration and further compression of the preexisting cosmic rays can explain the gamma-ray data. Within the hadronic scenario of the S147 gamma-ray origin, the derived gamma-ray luminosity would correspond to the total energy in CR protons $\sim$ 10$^{49}/n_a$ ergs (where $n_a$ which is measured in cm$^{-3}$ is the average number density of the gas in a single zone homogeneous model).

Later on, \citet{Suzuki22} obtained the 1-100 GeV Fermi-LAT gamma-ray luminosity of S147 
$\sim 6\times 10^{32}$ \ergs and the spectrum fitted with a broken power law model 
which has a photon index of about 2.14-2.18 below the break energy $\sim 1.4$ GeV and a much softer spectrum of a photon index $\sim 3.9$ above the break which extends down to the cutoff energies $> 21$ GeV. The radio spectrum discussed above is broadly consistent with that of the low energy gamma-rays having in mind that the momentum distributions of both electrons and protons accelerated by diffusive shock acceleration are expected to have the same power law index (of about 2) in the GeV energy range.   
The break in the gamma-ray spectrum at 1.4 GeV detected by Fermi-LAT within the hadronic model roughly corresponds to a break in the proton spectrum at about 14 GeV. Then assuming that the electron spectrum has a break at the same energy from the position of the break frequency in the radio spectrum given by \citet{2008A&A...482..783X} one can estimate the magnetic field in the diffuse emission to be about $\mu$G. If both the filaments and diffuse regions are embedded in a single population of DSA accelerated electrons then the lack of observed break in the radio filaments till 40 GHz implies the magnetic fields there to be above 40 $\mu$G.

\section{Discussion}
\label{sec:discussion}

Initially, S147 was recognized as a very old $\sim 10^5\,{\rm yr}$ SNR \cite[e.g.][]{1973ApJ...181..799S}. The discovery of the pulsar~J0538+2817 and the analysis of its kinematics suggested a significantly younger age $\sim 3-4\times 10^4\,{\rm yr}$   \citep{Romani_2003,Chatterjee_2009}, if the pulsar is indeed associated with S147. This duality in the age estimate translates into two possible scenarios: a very old SNR in a uniform medium or a younger SNR in a cavity.  The former scenario is discussed in a companion paper \citep{Miltos2024}, here we focus on the latter. Both papers make use of the new X-ray data provided by SRG/eROSITA.

Several generic scenarios of "a supernova in the cavity" have been modeled before \citep[e.g.][]{1990MNRAS.244..563T} and suggested for S147 \citep[e.g.][]{Reich_2003}. Here, to make the formulation of the model more complete, we assume that the cavity is created by the winds of a sufficiently massive star during its Main Sequence (MS) evolutionary phase.  The presence of the pulsar and the size of the ${\rm H}_\alpha$ nebula together suggest the progenitor mass of order $20 \, {\rm M_\odot}$, leading to a well-constrained model. It appears that this model can explain many X-ray properties of S147. In particular, this model naturally predicts X-ray spectra that feature lines of Mg and Si along with the lines of OVII and OVII, and does not require excess low-energy absorption. It also predicts that the X-ray emitting gas extends all the way to the  ${\rm H}_\alpha$-emitting shell. 

We note here, that a qualitatively similar picture is expected if the cavity, needed to reduce the SNR age, 
is produced by another mechanism rather than MS-driven WBB \citep[e.g][]{2006A&A...454..239G}. As long as the mass of the dense shell is large and the cavity is filled with low-density gas, most of the conclusions stay unchanged. From this point of view, the only important parameter for modeling the expected X-ray emission is the observed size of the SNR, which in this model is essentially the original size of the cavity.

A natural implication of this model is that the ions are much hotter in the interior of the dense shell than electrons and that the ionization equilibrium is not achieved. A direct test of these predictions should be possible with forthcoming X-ray bolometers like XRISM \citep[][]{2020arXiv200304962X}, ATHENA \citep[][]{2013arXiv1306.2307N}, and LEM \citep[][]{2022arXiv221109827K}. Detection of very broad lines of O, Ne, Mg, and Si ions would be a decisive test of the S147 "hot" scenario. The lack of broadening will instead support the "cold" model. For the time being, S147 can be considered a promising candidate for the list of objects featuring $T_i\gg T_e$ \citep[see, e.g.][]{2023ApJ...949...50R}. Constraints on weaker lines, e.g. Fe~XVII, from forthcoming bolometers would also tighten the constraints on the NEI models. 

As a caveat, we mention that it is not entirely clear if the modest abundance enhancement inferred from the observed flux of the NEI plasma represents a serious challenge to the hot model. Potentially, the abundance can be much higher.  Yet another simplification used here is the assumption of spherical symmetry, which clearly does not apply to S147.

Another interesting question is how often one could find SN~II exploded inside the WBB? 
The answer depends on the progenitor velocity ($v_{\ast}$) relative to the ISM. 
The dispersion velocity of OB-stars is $\sim 10$\kms, \citep{Bobylev_2022}, although well known RSG
Betelgeuse ($M\sim 15$\msun) shows an even larger velocity wrt ISM,  
$v_{\ast}\approx 30$\kms\ \citep{Ueta_2008}. 
The 20\msun\ star with the velocity of 10\kms\ wrt  ISM  will run 80\,pc to the end of the main sequence lifetime and will reside near the bubble frontal boundary at the distance comparable to the termination shock radius ($\sim 3$\,pc). 
At the subsequent helium burning stage ($0.8\times10^6$ yr) the star will run another 8\,pc and escape the bubble.
We conclude that most of SNe~II explode outside their WBB and S147 should be considered as a rare example of SN~II exploded in the WBB presumably owing to the low progenitor velocity wrt ISM.
This might explain a highly unusual H$\alpha$ appearance of S147, although Vela SNR seems to show a somewhat similar H$\alpha$ pattern.
Interestingly that a finite progenitor velocity of $\sim 1$\kms\ could results in the supernova offset wrt the bubble center \citep{Mackey_2015},
which in turn could bring about the observed asphericity of S147.

A somewhat disturbing fact, at first glance, is that we do not see any traces of a host stellar association, in which 20\msun\ star might be born. 
Indeed, it is commonly conjectured that all massive stars are born in OB associations or clusters.
However, this opinion is an oversimplification: a dedicated study of this issue for MW and LMC 
demonstrates that 5-10\% of massive stars are born in isolation 
\citep{Chu_2008}.

In the model discussed here, the age estimate based on the kinematics of the pulsar J0538+2817 broadly agrees with the observed characteristics of S147. This strengthens the association of the pulsar with the supernova. This implies that the pulsar is plausibly inside the SNR and might affect its properties. Some tantalizing hints might be spotted in Fig.\ref{fig:xrayrgb}. This will be the subject of a separate study.

To conclude this discussion, we reiterate that a more canonical scenario of an older SNR in a uniform medium is analyzed in detail in a companion paper of \cite{Miltos2024}.

\section{Conclusions}
\label{sec:conclusions}
We argue that new X-ray data of SRG/eROSITA lends support for S147 being a type II SN exploded in a low-density cavity. The main features of this scenario can be summarized as follows.

\begin{itemize}
\item The supernova went off in a cavity that was formed during MS stage of a $\sim 20 \,M_\odot $ star is a plausible scenario that leads to the formation of the cavity. 
\item This star had a relatively small velocity relative to the ambient ISM. As a result, a dense shell around the cavity is approximately spherical and the SN explosion was not far from its center.
\item The age is consistent with that derived from the kinematics of the pulsar PSR J0538+2817 ($\sim 35\,{\rm kyr}$).
\item The X-ray emission comes predominantly from the gas inside the cavity. The gas there has $T_e\ll T_i$ and is not in the collisional ionization equilibrium. This gas was reheated by reverse shock following the collision of the ejecta with the dense shell. A large fraction of ejecta kinetic energy is "reflected" by the shell back into the cavity. 
\item The ${\rm H_\alpha}$ and narrow radio filaments are associated with the radiative forward shock that propagates through the dense shell. 
\item Diffuse radio is due to the particles accelerated at the radiative shocks, but now living in an environment with a much weaker magnetic field.
\item Eventual fate of an SNR in the cavity:  hot and low-density gas bubble (no oxygen lines) until it dissolves in the ISM. 
\item Other scenarios for the cavity formation might be consistent with X-ray data, as long as the mass of the dense shell is large.
\end{itemize}

Finally, we note that S147 has a very complicated morphology and spectra from radio to gamma-rays and the above model is not intended to reproduce all this complexity. However, we believe that it captures some of the essential features of this remarkable SNR.

\section*{Acknowledgments}

This work is partly based on observations with the eROSITA telescope onboard \textit{SRG} space observatory. The \textit{SRG} observatory was built by Roskosmos in the interests of the Russian Academy of Sciences represented by its Space Research Institute (IKI) in the framework of the Russian Federal Space Program, with the participation of the Deutsches Zentrum für Luft- und Raumfahrt (DLR). The eROSITA X-ray telescope was built by a consortium of German Institutes led by MPE, and supported by DLR. The \textit{SRG} spacecraft was designed, built, launched, and is operated by the Lavochkin Association and its subcontractors. The science data are downlinked via the Deep Space Network Antennae in Bear Lakes, Ussurijsk, and Baikonur, funded by Roskosmos. 

The development and construction of the eROSITA X-ray instrument was led by MPE, with contributions from the Dr. Karl Remeis Observatory Bamberg $\&$ ECAP (FAU Erlangen-Nuernberg), the University of Hamburg Observatory, the Leibniz Institute for Astrophysics Potsdam (AIP), and the Institute for Astronomy and Astrophysics of the University of Tübingen, with the support of DLR and the Max Planck Society. The Argelander Institute for Astronomy of the University of Bonn and the Ludwig Maximilians Universität Munich also participated in the science preparation for eROSITA. The eROSITA data were processed using the eSASS/NRTA software system developed by the German eROSITA consortium and analysed using proprietary data reduction software developed by the Russian eROSITA Consortium.



IK acknowledges support by the COMPLEX project from the European Research Council (ERC) under the European Union’s Horizon 2020 research and innovation program grant agreement ERC-2019-AdG 882679. A.M.B. was supported by the RSF grant 21-72-20020. His modeling was performed at the Joint Supercomputer Center JSCC RAS and at the Peter the Great Saint-Petersburg Polytechnic University Supercomputing Center.

This research made use of \texttt{Montage}\footnote{\url{http://montage.ipac.caltech.edu}}. It is funded by the National Science Foundation under Grant Number ACI-1440620, and was previously funded by the National Aeronautics and Space Administration's Earth Science Technology Office, Computation Technologies Project, under Cooperative Agreement Number NCC5-626 between NASA and the California Institute of Technology.



\bibliographystyle{aa}
\bibliography{ref} 

\begin{thebibliography}{}
\makeatletter
\relax
\def\mn@urlcharsother{\let\do\@makeother \do\$\do\&\do\#\do\^\do\_\do\%\do\~}
\def\mn@doi{\begingroup\mn@urlcharsother \@ifnextchar [ {\mn@doi@}
  {\mn@doi@[]}}
\def\mn@doi@[#1]#2{\def\@tempa{#1}\ifx\@tempa\@empty \href
  {http://dx.doi.org/#2} {doi:#2}\else \href {http://dx.doi.org/#2} {#1}\fi
  \endgroup}
\def\mn@eprint#1#2{\mn@eprint@#1:#2::\@nil}
\def\mn@eprint@arXiv#1{\href {http://arxiv.org/abs/#1} {{\tt arXiv:#1}}}
\def\mn@eprint@dblp#1{\href {http://dblp.uni-trier.de/rec/bibtex/#1.xml}
  {dblp:#1}}
\def\mn@eprint@#1:#2:#3:#4\@nil{\def\@tempa {#1}\def\@tempb {#2}\def\@tempc
  {#3}\ifx \@tempc \@empty \let \@tempc \@tempb \let \@tempb \@tempa \fi \ifx
  \@tempb \@empty \def\@tempb {arXiv}\fi \@ifundefined
  {mn@eprint@\@tempb}{\@tempb:\@tempc}{\expandafter \expandafter \csname
  mn@eprint@\@tempb\endcsname \expandafter{\@tempc}}}

\bibitem[\protect\citeauthoryear{{Anderson}, {Cadwell}, {Jacoby}, {Wolszczan},
  {Foster}  \& {Kramer}}{{Anderson} et~al.}{1996}]{1996ApJ...468L..55A}
{Anderson} S.~B.,  {Cadwell} B.~J.,  {Jacoby} B.~A.,  {Wolszczan} A.,  {Foster}
  R.~S.,   {Kramer} M.,  1996, \mn@doi [\apjl] {10.1086/310218}, \href
  {https://ui.adsabs.harvard.edu/abs/1996ApJ...468L..55A} {468, L55}

\bibitem[\protect\citeauthoryear{{Arnaud}}{{Arnaud}}{1996}]{1996ASPC..101...17A}
{Arnaud} K.~A.,  1996, in {Jacoby} G.~H.,  {Barnes} J.,  eds,  Astronomical
  Society of the Pacific Conference Series Vol. 101, Astronomical Data Analysis
  Software and Systems V. p.~17

\bibitem[\protect\citeauthoryear{{Bertoldi} \& {McKee}}{{Bertoldi} \&
  {McKee}}{1990}]{Bertoldi_1990}
{Bertoldi} F.,  {McKee} C.~F.,  1990, \mn@doi [\apj] {10.1086/168713}, \href
  {https://ui.adsabs.harvard.edu/abs/1990ApJ...354..529B} {354, 529}

\bibitem[\protect\citeauthoryear{{Blondin}, {Wright}, {Borkowski}  \&
  {Reynolds}}{{Blondin} et~al.}{1998a}]{1998ApJ...500..342B}
{Blondin} J.~M.,  {Wright} E.~B.,  {Borkowski} K.~J.,   {Reynolds} S.~P.,
  1998a, \mn@doi [\apj] {10.1086/305708}, \href
  {https://ui.adsabs.harvard.edu/abs/1998ApJ...500..342B} {500, 342}

\bibitem[\protect\citeauthoryear{{Blondin}, {Wright}, {Borkowski}  \&
  {Reynolds}}{{Blondin} et~al.}{1998b}]{Blondin_1998}
{Blondin} J.~M.,  {Wright} E.~B.,  {Borkowski} K.~J.,   {Reynolds} S.~P.,
  1998b, \mn@doi [\apj] {10.1086/305708}, \href
  {https://ui.adsabs.harvard.edu/abs/1998ApJ...500..342B} {500, 342}

\bibitem[\protect\citeauthoryear{{Bobylev}, {Bajkova}  \& {Karelin}}{{Bobylev}
  et~al.}{2022}]{Bobylev_2022}
{Bobylev} V.~V.,  {Bajkova} A.~T.,   {Karelin} G.~M.,  2022, \mn@doi [Astronomy
  Letters] {10.1134/S1063773722040016}, \href
  {https://ui.adsabs.harvard.edu/abs/2022AstL...48..243B} {48, 243}

\bibitem[\protect\citeauthoryear{{Bykov}, {Ellison}, {Marcowith}  \&
  {Osipov}}{{Bykov} et~al.}{2018}]{CR_B18}
{Bykov} A.~M.,  {Ellison} D.~C.,  {Marcowith} A.,   {Osipov} S.~M.,  2018,
  \mn@doi [\ssr] {10.1007/s11214-018-0479-4}, \href
  {https://ui.adsabs.harvard.edu/abs/2018SSRv..214...41B} {214, 41}

\bibitem[\protect\citeauthoryear{{Chatterjee} et~al.,}{{Chatterjee}
  et~al.}{2009}]{Chatterjee_2009}
{Chatterjee} S.,  et~al., 2009, \mn@doi [\apj] {10.1088/0004-637X/698/1/250},
  \href {https://ui.adsabs.harvard.edu/abs/2009ApJ...698..250C} {698, 250}

\bibitem[\protect\citeauthoryear{{Chevalier} \& {Liang}}{{Chevalier} \&
  {Liang}}{1989}]{1989ApJ...344..332C}
{Chevalier} R.~A.,  {Liang} E.~P.,  1989, \mn@doi [\apj] {10.1086/167802},
  \href {https://ui.adsabs.harvard.edu/abs/1989ApJ...344..332C} {344, 332}

\bibitem[\protect\citeauthoryear{{Chu} \& {Gruendl}}{{Chu} \&
  {Gruendl}}{2008}]{Chu_2008}
{Chu} Y.~H.,  {Gruendl} R.~A.,  2008, in {Beuther} H.,  {Linz} H.,   {Henning}
  T.,  eds,  Astronomical Society of the Pacific Conference Series Vol. 387,
  Massive Star Formation: Observations Confront Theory. p.~415 (\mn@eprint
  {arXiv} {0712.1871}), \mn@doi{10.48550/arXiv.0712.1871}

\bibitem[\protect\citeauthoryear{{Churazov}, {Khabibullin}, {Bykov}, {Chugai},
  {Sunyaev}  \& {Zinchenko}}{{Churazov} et~al.}{2021}]{2021MNRAS.507..971C}
{Churazov} E.~M.,  {Khabibullin} I.~I.,  {Bykov} A.~M.,  {Chugai} N.~N.,
  {Sunyaev} R.~A.,   {Zinchenko} I.~I.,  2021, \mn@doi [\mnras]
  {10.1093/mnras/stab2125}, \href
  {https://ui.adsabs.harvard.edu/abs/2021MNRAS.507..971C} {507, 971}

\bibitem[\protect\citeauthoryear{{Cox}}{{Cox}}{2005}]{Cox_2005}
{Cox} D.~P.,  2005, \mn@doi [\araa] {10.1146/annurev.astro.43.072103.150615},
  \href {https://ui.adsabs.harvard.edu/abs/2005ARA&A..43..337C} {43, 337}

\bibitem[\protect\citeauthoryear{{Denoyer}}{{Denoyer}}{1974}]{1974AJ.....79.1253D}
{Denoyer} L.~K.,  1974, \mn@doi [\aj] {10.1086/111668}, \href
  {https://ui.adsabs.harvard.edu/abs/1974AJ.....79.1253D} {79, 1253}

\bibitem[\protect\citeauthoryear{{Dickey} \& {Garwood}}{{Dickey} \&
  {Garwood}}{1989}]{Dickey_1989}
{Dickey} J.~M.,  {Garwood} R.~W.,  1989, \mn@doi [\apj] {10.1086/167485}, \href
  {https://ui.adsabs.harvard.edu/abs/1989ApJ...341..201D} {341, 201}

\bibitem[\protect\citeauthoryear{{Dwarkadas}}{{Dwarkadas}}{2023}]{2023Galax..11...78D}
{Dwarkadas} V.~V.,  2023, \mn@doi [Galaxies] {10.3390/galaxies11030078}, \href
  {https://ui.adsabs.harvard.edu/abs/2023Galax..11...78D} {11, 78}

\bibitem[\protect\citeauthoryear{{Fuerst} \& {Reich}}{{Fuerst} \&
  {Reich}}{1986}]{1986A&A...163..185F}
{Fuerst} E.,  {Reich} W.,  1986, \aap, \href
  {https://ui.adsabs.harvard.edu/abs/1986A&A...163..185F} {163, 185}

\bibitem[\protect\citeauthoryear{{Gaze} \& {Shajn}}{{Gaze} \&
  {Shajn}}{1952}]{1952IzKry...9...52G}
{Gaze} V.~F.,  {Shajn} G.~A.,  1952, Izvestiya Krymskoj Astrofizicheskoj
  Observatorii, \href {https://ui.adsabs.harvard.edu/abs/1952IzKry...9...52G}
  {9, 52}

\bibitem[\protect\citeauthoryear{{Greimel} et~al.,}{{Greimel}
  et~al.}{2021}]{2021A&A...655A..49G}
{Greimel} R.,  et~al., 2021, \mn@doi [\aap] {10.1051/0004-6361/202140950},
  \href {https://ui.adsabs.harvard.edu/abs/2021A&A...655A..49G} {655, A49}

\bibitem[\protect\citeauthoryear{{Gvaramadze}}{{Gvaramadze}}{2006}]{2006A&A...454..239G}
{Gvaramadze} V.~V.,  2006, \mn@doi [\aap] {10.1051/0004-6361:20054114}, \href
  {https://ui.adsabs.harvard.edu/abs/2006A&A...454..239G} {454, 239}

\bibitem[\protect\citeauthoryear{{Heger}, {Fryer}, {Woosley}, {Langer}  \&
  {Hartmann}}{{Heger} et~al.}{2003}]{Heger_2003}
{Heger} A.,  {Fryer} C.~L.,  {Woosley} S.~E.,  {Langer} N.,   {Hartmann} D.~H.,
   2003, \mn@doi [\apj] {10.1086/375341}, \href
  {http://adsabs.harvard.edu/abs/2003ApJ...591..288H} {591, 288}

\bibitem[\protect\citeauthoryear{{Heyer} \& {Dame}}{{Heyer} \&
  {Dame}}{2015}]{Heyer2015}
{Heyer} M.,  {Dame} T.~M.,  2015, \mn@doi [\araa]
  {10.1146/annurev-astro-082214-122324}, \href
  {https://ui.adsabs.harvard.edu/abs/2015ARA&A..53..583H} {53, 583}

\bibitem[\protect\citeauthoryear{{Howarth} \& {Prinja}}{{Howarth} \&
  {Prinja}}{1989}]{Howarth_1989}
{Howarth} I.~D.,  {Prinja} R.~K.,  1989, \mn@doi [\apjs] {10.1086/191321},
  \href {https://ui.adsabs.harvard.edu/abs/1989ApJS...69..527H} {69, 527}

\bibitem[\protect\citeauthoryear{{Hughes}}{{Hughes}}{1987}]{Hughes_1987}
{Hughes} J.~P.,  1987, \mn@doi [\apj] {10.1086/165043}, \href
  {https://ui.adsabs.harvard.edu/abs/1987ApJ...314..103H} {314, 103}

\bibitem[\protect\citeauthoryear{{Katsuta} et~al.,}{{Katsuta}
  et~al.}{2012}]{2012ApJ...752..135K}
{Katsuta} J.,  et~al., 2012, \mn@doi [\apj] {10.1088/0004-637X/752/2/135},
  \href {https://ui.adsabs.harvard.edu/abs/2012ApJ...752..135K} {752, 135}

\bibitem[\protect\citeauthoryear{{Khabibullin}, {Churazov}, {Bykov}, {Chugai}
  \& {Sunyaev}}{{Khabibullin} et~al.}{2023}]{2023MNRAS.521.5536K}
{Khabibullin} I.~I.,  {Churazov} E.~M.,  {Bykov} A.~M.,  {Chugai} N.~N.,
  {Sunyaev} R.~A.,  2023, \mn@doi [\mnras] {10.1093/mnras/stad818}, \href
  {https://ui.adsabs.harvard.edu/abs/2023MNRAS.521.5536K} {521, 5536}

\bibitem[\protect\citeauthoryear{{Khabibullin}, {Churazov}, {Bykov}, {Chugai}
  \& {Zinchenko}}{{Khabibullin} et~al.}{2024}]{2024MNRAS.527.5683K}
{Khabibullin} I.~I.,  {Churazov} E.~M.,  {Bykov} A.~M.,  {Chugai} N.~N.,
  {Zinchenko} I.~I.,  2024, \mn@doi [\mnras] {10.1093/mnras/stad3452}, \href
  {https://ui.adsabs.harvard.edu/abs/2024MNRAS.527.5683K} {527, 5683}

\bibitem[\protect\citeauthoryear{{Kirshner} \& {Arnold}}{{Kirshner} \&
  {Arnold}}{1979}]{Kirshner_1979}
{Kirshner} R.~P.,  {Arnold} C.~N.,  1979, \mn@doi [\apj] {10.1086/156939},
  \href {https://ui.adsabs.harvard.edu/abs/1979ApJ...229..147K} {229, 147}

\bibitem[\protect\citeauthoryear{{Kraft} et~al.,}{{Kraft}
  et~al.}{2022}]{2022arXiv221109827K}
{Kraft} R.,  et~al., 2022, \mn@doi [arXiv e-prints]
  {10.48550/arXiv.2211.09827}, \href
  {https://ui.adsabs.harvard.edu/abs/2022arXiv221109827K} {p. arXiv:2211.09827}

\bibitem[\protect\citeauthoryear{{Kramer}, {Lyne}, {Hobbs}, {L{\"o}hmer},
  {Carr}, {Jordan}  \& {Wolszczan}}{{Kramer}
  et~al.}{2003}]{2003ApJ...593L..31K}
{Kramer} M.,  {Lyne} A.~G.,  {Hobbs} G.,  {L{\"o}hmer} O.,  {Carr} P.,
  {Jordan} C.,   {Wolszczan} A.,  2003, \mn@doi [\apjl] {10.1086/378082}, \href
  {https://ui.adsabs.harvard.edu/abs/2003ApJ...593L..31K} {593, L31}

\bibitem[\protect\citeauthoryear{{Larson}}{{Larson}}{1981}]{Larson1981}
{Larson} R.~B.,  1981, \mn@doi [\mnras] {10.1093/mnras/194.4.809}, \href
  {https://ui.adsabs.harvard.edu/abs/1981MNRAS.194..809L} {194, 809}

\bibitem[\protect\citeauthoryear{{Liedahl}, {Osterheld}  \&
  {Goldstein}}{{Liedahl} et~al.}{1995}]{1995ApJ...438L.115L}
{Liedahl} D.~A.,  {Osterheld} A.~L.,   {Goldstein} W.~H.,  1995, \mn@doi
  [\apjl] {10.1086/187729}, \href
  {https://ui.adsabs.harvard.edu/abs/1995ApJ...438L.115L} {438, L115}

\bibitem[\protect\citeauthoryear{{Lozinskaia}}{{Lozinskaia}}{1976}]{1976AZh....53...38L}
{Lozinskaia} T.~A.,  1976, \azh, \href
  {https://ui.adsabs.harvard.edu/abs/1976AZh....53...38L} {53, 38}

\bibitem[\protect\citeauthoryear{{Lozinskaya}}{{Lozinskaya}}{1976}]{Lozinskaya_1976}
{Lozinskaya} T.~A.,  1976, \sovast, \href
  {https://ui.adsabs.harvard.edu/abs/1976SvA....20...19L} {20, 19}

\bibitem[\protect\citeauthoryear{{Mackey}, {Gvaramadze}, {Mohamed}  \&
  {Langer}}{{Mackey} et~al.}{2015}]{Mackey_2015}
{Mackey} J.,  {Gvaramadze} V.~V.,  {Mohamed} S.,   {Langer} N.,  2015, \mn@doi
  [\aap] {10.1051/0004-6361/201424716}, \href
  {https://ui.adsabs.harvard.edu/abs/2015A&A...573A..10M} {573, A10}

\bibitem[\protect\citeauthoryear{{Martizzi}, {Faucher-Gigu{\`e}re}  \&
  {Quataert}}{{Martizzi} et~al.}{2015}]{Martizzi_2015}
{Martizzi} D.,  {Faucher-Gigu{\`e}re} C.-A.,   {Quataert} E.,  2015, \mn@doi
  [\mnras] {10.1093/mnras/stv562}, \href
  {https://ui.adsabs.harvard.edu/abs/2015MNRAS.450..504M} {450, 504}

\bibitem[\protect\citeauthoryear{{Mewe}, {Gronenschild}  \& {van den
  Oord}}{{Mewe} et~al.}{1985}]{1985A&AS...62..197M}
{Mewe} R.,  {Gronenschild} E.~H.~B.~M.,   {van den Oord} G.~H.~J.,  1985,
  \aaps, \href {https://ui.adsabs.harvard.edu/abs/1985A&AS...62..197M} {62,
  197}

\bibitem[\protect\citeauthoryear{{Michailidis et al.}}{{Michailidis et
  al.}}{2024}]{Miltos2024}
{Michailidis et al.} 2024, \aap

\bibitem[\protect\citeauthoryear{{Nandra} et~al.,}{{Nandra}
  et~al.}{2013}]{2013arXiv1306.2307N}
{Nandra} K.,  et~al., 2013, \mn@doi [arXiv e-prints]
  {10.48550/arXiv.1306.2307}, \href
  {https://ui.adsabs.harvard.edu/abs/2013arXiv1306.2307N} {p. arXiv:1306.2307}

\bibitem[\protect\citeauthoryear{{Ng}, {Romani}, {Brisken}, {Chatterjee}  \&
  {Kramer}}{{Ng} et~al.}{2007}]{2007ApJ...654..487N}
{Ng} C.~Y.,  {Romani} R.~W.,  {Brisken} W.~F.,  {Chatterjee} S.,   {Kramer} M.,
   2007, \mn@doi [\apj] {10.1086/510576}, \href
  {https://ui.adsabs.harvard.edu/abs/2007ApJ...654..487N} {654, 487}

\bibitem[\protect\citeauthoryear{{Pavlinsky} et~al.,}{{Pavlinsky}
  et~al.}{2021}]{2021A&A...650A..42P}
{Pavlinsky} M.,  et~al., 2021, \mn@doi [\aap] {10.1051/0004-6361/202040265},
  \href {https://ui.adsabs.harvard.edu/abs/2021A&A...650A..42P} {650, A42}

\bibitem[\protect\citeauthoryear{{Pikel'ner}}{{Pikel'ner}}{1954}]{Pikelner_1954}
{Pikel'ner} S.,  1954, Izvestiya Ordena Trudovogo Krasnogo Znameni Krymskoj
  Astrofizicheskoj Observatorii, \href
  {https://ui.adsabs.harvard.edu/abs/1954IzKry..12...93P} {12, 93}

\bibitem[\protect\citeauthoryear{{Predehl} et~al.,}{{Predehl}
  et~al.}{2021}]{2021A&A...647A...1P}
{Predehl} P.,  et~al., 2021, \mn@doi [\aap] {10.1051/0004-6361/202039313},
  \href {https://ui.adsabs.harvard.edu/abs/2021A&A...647A...1P} {647, A1}

\bibitem[\protect\citeauthoryear{{Raymond} et~al.,}{{Raymond}
  et~al.}{2023}]{2023ApJ...949...50R}
{Raymond} J.~C.,  et~al., 2023, \mn@doi [\apj] {10.3847/1538-4357/acc528},
  \href {https://ui.adsabs.harvard.edu/abs/2023ApJ...949...50R} {949, 50}

\bibitem[\protect\citeauthoryear{{Reich}, {Zhang}  \& {F{\"u}rst}}{{Reich}
  et~al.}{2003}]{Reich_2003}
{Reich} W.,  {Zhang} X.,   {F{\"u}rst} E.,  2003, \mn@doi [\aap]
  {10.1051/0004-6361:20030939}, \href
  {https://ui.adsabs.harvard.edu/abs/2003A&A...408..961R} {408, 961}

\bibitem[\protect\citeauthoryear{{Ren} et~al.,}{{Ren} et~al.}{2018}]{Ren_2018}
{Ren} J.-J.,  et~al., 2018, \mn@doi [Research in Astronomy and Astrophysics]
  {10.1088/1674-4527/18/9/111}, \href
  {https://ui.adsabs.harvard.edu/abs/2018RAA....18..111R} {18, 111}

\bibitem[\protect\citeauthoryear{{Romani} \& {Ng}}{{Romani} \&
  {Ng}}{2003}]{Romani_2003}
{Romani} R.~W.,  {Ng} C.~Y.,  2003, \mn@doi [\apjl] {10.1086/374259}, \href
  {https://ui.adsabs.harvard.edu/abs/2003ApJ...585L..41R} {585, L41}

\bibitem[\protect\citeauthoryear{{Schaller}, {Schaerer}, {Meynet}  \&
  {Maeder}}{{Schaller} et~al.}{1992}]{Schaller_1992}
{Schaller} G.,  {Schaerer} D.,  {Meynet} G.,   {Maeder} A.,  1992, \aaps, \href
  {https://ui.adsabs.harvard.edu/abs/1992A&AS...96..269S} {96, 269}

\bibitem[\protect\citeauthoryear{{Silk} \& {Wallerstein}}{{Silk} \&
  {Wallerstein}}{1973}]{1973ApJ...181..799S}
{Silk} J.,  {Wallerstein} G.,  1973, \mn@doi [\apj] {10.1086/152090}, \href
  {https://ui.adsabs.harvard.edu/abs/1973ApJ...181..799S} {181, 799}

\bibitem[\protect\citeauthoryear{{Sofue}, {Furst}  \& {Hirth}}{{Sofue}
  et~al.}{1980}]{1980PASJ...32....1S}
{Sofue} Y.,  {Furst} E.,   {Hirth} W.,  1980, \pasj, \href
  {https://ui.adsabs.harvard.edu/abs/1980PASJ...32....1S} {32, 1}

\bibitem[\protect\citeauthoryear{{Sunyaev} et~al.,}{{Sunyaev}
  et~al.}{2021}]{2021A&A...656A.132S}
{Sunyaev} R.,  et~al., 2021, \mn@doi [\aap] {10.1051/0004-6361/202141179},
  \href {https://ui.adsabs.harvard.edu/abs/2021A&A...656A.132S} {656, A132}

\bibitem[\protect\citeauthoryear{{Sutherland} \& {Dopita}}{{Sutherland} \&
  {Dopita}}{1993}]{Sutherland_1993}
{Sutherland} R.~S.,  {Dopita} M.~A.,  1993, \mn@doi [\apjs] {10.1086/191823},
  \href {https://ui.adsabs.harvard.edu/abs/1993ApJS...88..253S} {88, 253}

\bibitem[\protect\citeauthoryear{{Suzuki}, {Bamba}, {Yamazaki}  \&
  {Ohira}}{{Suzuki} et~al.}{2022}]{Suzuki22}
{Suzuki} H.,  {Bamba} A.,  {Yamazaki} R.,   {Ohira} Y.,  2022, \mn@doi [\apj]
  {10.3847/1538-4357/ac33b5}, \href
  {https://ui.adsabs.harvard.edu/abs/2022ApJ...924...45S} {924, 45}

\bibitem[\protect\citeauthoryear{{Taylor} et~al.,}{{Taylor}
  et~al.}{2003}]{2003AJ....125.3145T}
{Taylor} A.~R.,  et~al., 2003, \mn@doi [\aj] {10.1086/375301}, \href
  {https://ui.adsabs.harvard.edu/abs/2003AJ....125.3145T} {125, 3145}

\bibitem[\protect\citeauthoryear{{Tenorio-Tagle}, {Bodenheimer}, {Franco}  \&
  {Rozyczka}}{{Tenorio-Tagle} et~al.}{1990}]{1990MNRAS.244..563T}
{Tenorio-Tagle} G.,  {Bodenheimer} P.,  {Franco} J.,   {Rozyczka} M.,  1990,
  \mnras, \href {https://ui.adsabs.harvard.edu/abs/1990MNRAS.244..563T} {244,
  563}

\bibitem[\protect\citeauthoryear{{Tenorio-Tagle}, {Rozyczka}, {Franco}  \&
  {Bodenheimer}}{{Tenorio-Tagle} et~al.}{1991}]{1991MNRAS.251..318T}
{Tenorio-Tagle} G.,  {Rozyczka} M.,  {Franco} J.,   {Bodenheimer} P.,  1991,
  \mn@doi [\mnras] {10.1093/mnras/251.2.318}, \href
  {https://ui.adsabs.harvard.edu/abs/1991MNRAS.251..318T} {251, 318}

\bibitem[\protect\citeauthoryear{{Ueta} et~al.,}{{Ueta}
  et~al.}{2008}]{Ueta_2008}
{Ueta} T.,  et~al., 2008, \mn@doi [\pasj] {10.1093/pasj/60.sp2.S407}, \href
  {https://ui.adsabs.harvard.edu/abs/2008PASJ...60S.407U} {60, S407}

\bibitem[\protect\citeauthoryear{{Utrobin}, {Wongwathanarat}, {Janka}  \&
  {M{\"u}ller}}{{Utrobin} et~al.}{2017}]{Utrobin_17}
{Utrobin} V.~P.,  {Wongwathanarat} A.,  {Janka} H.-T.,   {M{\"u}ller} E.,
  2017, \mn@doi [\apj] {10.3847/1538-4357/aa8594}, \href
  {http://adsabs.harvard.edu/abs/2017ApJ...846...37U} {846, 37}

\bibitem[\protect\citeauthoryear{{Vink}}{{Vink}}{2020}]{vink20}
{Vink} J.,  2020, {Physics and Evolution of Supernova Remnants}.
Springer Nature, \mn@doi{10.1007/978-3-030-55231-2}

\bibitem[\protect\citeauthoryear{{Weaver}, {McCray}, {Castor}, {Shapiro}  \&
  {Moore}}{{Weaver} et~al.}{1977}]{Weaver_1977}
{Weaver} R.,  {McCray} R.,  {Castor} J.,  {Shapiro} P.,   {Moore} R.,  1977,
  \mn@doi [\apj] {10.1086/155692}, \href
  {https://ui.adsabs.harvard.edu/abs/1977ApJ...218..377W} {218, 377}

\bibitem[\protect\citeauthoryear{{Woosley}, {Heger}  \& {Weaver}}{{Woosley}
  et~al.}{2002}]{Woosley_2002}
{Woosley} S.~E.,  {Heger} A.,   {Weaver} T.~A.,  2002, \mn@doi [Reviews of
  Modern Physics] {10.1103/RevModPhys.74.1015}, \href
  {http://adsabs.harvard.edu/abs/2002RvMP...74.1015W} {74, 1015}

\bibitem[\protect\citeauthoryear{{XRISM Science Team}}{{XRISM Science
  Team}}{2020}]{2020arXiv200304962X}
{XRISM Science Team} 2020, arXiv e-prints, p. arXiv:2003.04962

\bibitem[\protect\citeauthoryear{{Xiao}, {F{\"u}rst}, {Reich}  \& {Han}}{{Xiao}
  et~al.}{2008}]{2008A&A...482..783X}
{Xiao} L.,  {F{\"u}rst} E.,  {Reich} W.,   {Han} J.~L.,  2008, \mn@doi [\aap]
  {10.1051/0004-6361:20078461}, \href
  {https://ui.adsabs.harvard.edu/abs/2008A&A...482..783X} {482, 783}

\makeatother
\end{thebibliography}





\label{lastpage}
\listofobjects

\end{document}